\documentclass[arial,11pt,a4paper]{article}

\pdfoutput=1 
\usepackage[left=3cm, right=3cm, top=3cm, bottom=3cm]{geometry}
\usepackage{graphicx,grffile}
\usepackage{amsmath,amsfonts,amssymb}
\usepackage[latin1]{inputenc}
\usepackage{newtxtext}
\usepackage{newtxmath}
\usepackage{natbib}
\usepackage{hyperref}
\hypersetup{
    colorlinks = true,
    urlcolor   = blue,
    citecolor  = black,
    linkcolor=blue,
}

\newcommand{\RomanNumeralCaps}[1]
\linenumbers
\usepackage{etoolbox}
\usepackage{tikz}
\usepackage{pgfplots}
\pgfplotsset{compat=1.14}
\usepgfplotslibrary{colormaps}
\usepackage{floatrow}
\usepackage{subfig}
\newrobustcmd*{\mysquare}[1]{\tikz{\draw[draw=#1, line width=1.2pt] (0,0) rectangle (0.15cm,0.15cm);}}
\newrobustcmd*{\mycircle}[1]{\tikz{\draw[draw=#1, line width=1.2pt] (0,0) circle [radius=0.075cm];}}
\newrobustcmd*{\mytriangle}[1]{\tikz{\draw[draw=#1, line width=1.2pt] (0,0) -- (0.15cm,0) -- (0.075cm,0.15cm) --(0,0);}}
\newrobustcmd*{\myline}[1]{\raisebox{2pt}{\tikz{\draw[draw=#1, line width = 1pt](0,0.8mm) -- (5.5mm,0.8mm)}}}
\newrobustcmd*{\mylinsym}[1]{\raisebox{-2pt}{\tikz{\draw [black, solid, line width=1pt] (0,.1mm) -- (5.5mm,.1mm) node [pos=.5] {#1};}}}
\usepackage{authblk}
\title{Direct numerical simulation of turbulent mass transfer at the surface of an open channel flow}
\author[1]{Michele Pinelli}
\author[1,*]{H. Herlina}
\author[2]{J.G. Wissink}
\author[1]{M. Uhlmann}
\affil[1]{Institute for Hydromechanics, Karlsruhe Institute of Technology, Kaiserstr.12, 76131 Karlsruhe, Germany}
\affil[2]{Department of Mechanical and Aerospace Engineering, Brunel University London, Kingston Lane, Uxbridge, UB8 3PH, UK}
\affil[*]{Corresponding author: herlina.herlina@kit.edu}
\date{}                     % if you don't need date to appear
\usepackage{fancyhdr}
\fancypagestyle{firstpage}{
  \lhead{\textit{Submitted to Journal of Fluid Mechanics
  \newline This is a pre-revision version
  \newline (revised version accepted for publication 20-Nov-2021)}}
}

\begin{document}
\maketitle
\thispagestyle{firstpage}

\begin{abstract}
We present direct numerical simulation results of turbulent open
channel flow at bulk Reynolds numbers up to 12000, coupled with
(passive) scalar transport at Schmidt numbers up to 200.
Care is taken to capture the very large scale motions which appear
already for relatively modest Reynolds numbers. 
The transfer velocity at the flat, free surface is found to scale with the Schmidt number to the power "-1/2", 
in accordance with previous studies and theoretical predictions 
for uncontaminated surfaces. 
The scaling of the transfer velocity with Reynolds number is found to vary, depending 
on the Reynolds number definition used. 
To compare the present results with
those obtained in other systems, we define a turbulent Reynolds number
at the edge of the surface-influenced layer. This allows us to probe
the two-regime model of Theofanous [Turbulent mass transfer at free,
gas-liquid interfaces, with applications to open-channel, bubble and
jet flows. Int. J. Heat Mass Transfer 19, 613--624, 1976], which is
found to correctly predict that small-scale vortices significantly affect the mass transfer for turbulent Reynolds numbers larger than 500. 
It is further established that the root-mean-square of the surface divergence is, on average,
proportional to the mean transfer velocity. However, the spatial correlation between instantaneous surface divergence and transfer velocity tends to decrease with increasing Schmidt number and increase with increasing Reynolds number. 
The latter is shown to be caused 
by an enhancement of the correlation in high-speed regions, which in turn is linked to the
spatial distribution of surface-parallel vortices.
\end{abstract}

%-------------------------------------------------------------------------
\section{Introduction}
%-------------------------------------------------------------------------
Transfer of gases across a gas-liquid interface is of fundamental importance in many research fields, such as civil engineering and environmental science. 
The transfer rate is governed by a variety of processes, the most important being the hydrodynamic flow conditions at the liquid side, which are significantly affected by the surface condition.
The latter includes surface contamination effects \citep[e.g.][]{tsai2003, McKenna2004, Wissink2017}, surface roughness effects \citep[e.g.][]{turney2016}  and whitecap effects  \citep[e.g.][]{jahne2020, brumer2017}.  
In nature, turbulent flow typically generated by wind-shear, buoyancy and/or bottom-shear enhances the gas transfer rate.  Wind-shear is usually the most effective driving mechanism and has been extensively studied \citep[e.g.][]{Plate84,komori93exp,Wanninkhof2009, garbe14}. At low wind-speeds, buoyancy and bottom-shear induced turbulence become more important. Particularly in deep water bodies, buoyancy effects due to surface cooling (e.g. by evaporation) often contribute significantly to the interfacial transfer of heat and greenhouse gases \citep[e.g.][]{podgrajsek14}. In rivers or streams, bottom-shear induced turbulence is generally the dominant gas transfer mechanism, which we aim to study here in detail. To this end, highly accurate direct numerical simulations (DNS) of mass transfer across a shear-free surface in isothermal open channel flow were performed. 

Commonly, the mass transfer across the air-water interface is measured using the transfer velocity $K_L$, which remains difficult to predict accurately. Firstly, because of the complex interrelated physical processes (including molecular diffusion, turbulent mixing, waves and surfactants), and secondly, due to the fact that the interfacial transfer processes of low (O$_2$, CO, CH$_4$) to moderate (CO$_2$) soluble gases are concentrated in a very thin boundary layer on the water side adjacent to the surface \citep{Jaehne1998}. 
Thus, traditional empirical models for the prediction of gas transfer in streams and rivers are mostly based on global hydraulic parameters, such as water depth, bulk velocity and bottom slope \citep[e.g.][]{Thackston1969,Gulliver1989}  or more specific parameters, such as  bed roughness  \citep[e.g.][]{Moog02}. 

Apart from empirical models, also several conceptual models have been proposed. An overview of several important conceptual models, such as the surface renewal model \citep{Danckwerts1951}, the large-eddy model \citep{Fortescue1967}, the small-eddy model \citep{Banerjee1968, Lamont1970}, and the surface divergence model \citep{McCready1986} can be found in e.g. \cite{theofanous:84,herlina:14}. 
Field measurements in various systems (including the coastal ocean and the large tidal freshwater river) by \cite{zappa2007} support the small-eddy model. By performing experiments in open channels, \cite{Plate84} and \cite{Moog1999} showed the applicability of the small-eddy model to predict interfacial mass transfer. 
More recently, \cite{Turney2013} experimentally validated the surface divergence model and showed that it can accurately predict open channel mass transfer in windless conditions. 

The very thin concentration boundary layer mentioned above makes it very difficult to perform measurements \citep[][]{herlina:08} as well as fully-resolved numerical simulations at realistic Reynolds and Prandtl / Schmidt numbers ($Pr=\nu/\kappa_h$, $Sc=\nu/D$, where  $\nu$  is the kinematic viscosity, $\kappa_h$ is the thermal diffusivity and $D$ is the molecular diffusivity). The main problem here is the very low diffusivity (high $Sc$) of many atmospheric gases in water, resulting in filaments with very large concentration gradients that need to be fully resolved. For this reason, DNS were usually performed at low to moderate Schmidt and/or Reynolds numbers. For example, \cite{lakehal2003} and \cite{tsai2013} performed DNS of interfacial mass transfer driven by surface-shear/wind for $Sc$ up to $10$ and $7$, respectively, while \cite{Schwertfirm2007} performed DNS  of buoyancy-driven mass transfer for $Sc$ up to $49$. Recently, DNS of high-realistic Schmidt numbers (e.g. $Sc \approx 500 $ for O$_2$ and $Sc\approx600$ for CO$_2$ in water) have been performed in buoyancy-driven flow by \cite{wissink:16,  fredriksson:16} and in isotropic-turbulence driven flow by \cite{herlina:14, herlina:19}. Note that the latter  is often used to produce a near-surface turbulent flow field that, despite the lack of streamwise anisotropy, is used to mimic the flow field generated by  bottom-shear, such as in open channel flow.

\cite{handler:99} were among the first to perform DNS of interfacial mass transfer in open channel flow, using a Prandtl number of $Pr=2$ and a friction Reynolds number of $Re_\tau=u_\tau H/\nu= 180$, where $u_\tau$ is the friction velocity and $H$ is the channel height. 
 It was observed that hairpin vortices were the dominant structures contributing to transfer of passive scalars in open channel flow. 
\cite{Nagaosa2003}, who performed DNS for $Re_\tau = 150, 300$ and $Pr=1$, showed that these hairpin vortices, which were generated  at the bottom wall, evolved into ring-like vortices as they approach the free-surface. These ring-like vortices were observed to be present immediately below high mass transfer regions. 
\cite{Kermani2011} performed DNS at $Re_\tau\simeq 300$ and $0.71\le Sc \le 8$. They focused on the quantification of the surface age (i.e. the time between two surface renewal events) and concluded that the random surface renewal model of \cite{Danckwerts1951} is not applicable at small (young) surface age, where interfacial mass transfer is actually at its largest. 
More recently, \cite{Nagaosa2012} performed DNS in open channel flow for $150\le Re_\tau\le 600$ with $Sc=1$. They evaluated the suitability of two characteristic time scales to approximate the renewal time in Danckwert's model. The first was based on the characteristic length and velocity scales at the free-surface, while the other was the reciprocal of the root-mean-square of the surface divergence. Both time scales were found to perform well.  
\cite{Magnaudet2006} used the computationally less demanding large eddy simulation (which only resolves the larger scales of motion) to be able to study interfacial mass transfer in open channel flow at a high friction Reynolds number of $Re_\tau=1280$ and Schmidt numbers ranging from $Sc=1$ to $200$. They confirmed the scaling of $K_L$ with $Sc^{-0.5}$ and found that the mass transfer was mostly driven by motions with a large streamwise extent.

In summary, because of the huge computational demands to fully resolve all scales of motion, until recently DNS of mass transfer in open channel flow was limited to  $Sc\leq8$ and $Re_\tau\leq600$, and relatively small computational domains. The latter making it impossible to capture the full extent of very large streamwise motions (VLSM).  
In this paper we present a detailed study of interfacial mass transfer for $180 \leq Re_\tau \leq 630$ in combination with  $Sc$ up to $200$ using computational domains sufficiently large to capture the VLSM. The main objectives are (i) to study the effect of anisotropy on interfacial mass transfer, (ii) to confirm the scaling of the transfer velocity with $Sc$  also for larger Schmidt numbers, (iii) to show that turbulence dissipation is an important factor in the dynamical behaviour of surface divergence, (iv) to evaluate the validity of Theofanous' model for open channel flow, and finally (v) to produce detailed data on near surface scalar statistics.

%-------------------------------------------------------------------------
\section{Numerical aspects}
%-------------------------------------------------------------------------
\subsection{Numerical method}
The mathematical formulation describing the process of mass transfer (e.g. dissolved gas transfer) in turbulent open channel flow comprises the incompressible Navier-Stokes equations for the flow and advection-diffusion equations for the scalar concentration field, which in non-dimensional form read
\begin{equation}
   \dfrac{\partial u^*_i}{\partial x^*_i} =  0,
   \label{eqn:cont}
\end{equation}
\begin{equation}
   \dfrac{\partial {u^*_i}}{\partial t^*}+\dfrac{\partial u^*_i u^*_j}{\partial x^*_j}
       = -\dfrac{\partial p^*}{\partial x^*_i}  
       +\dfrac{1}{Re_b} \dfrac{\partial ^2 u^*_i}{\partial x^*_j \partial x^*_j}
       + f^*\delta_{i1},
   \label{eqn:NS}
\end{equation}
\begin{equation}   
   \dfrac{\partial c^*}{\partial t^*}+\dfrac{\partial u^*_jc^*}{\partial x^*_j}     
      =\dfrac{1}{Re_b\,Sc} \dfrac{\partial ^2 c^*}{\partial x^*_j \partial x^*_j},
   \label{eqn:cg}
\end{equation}
\noindent where ${\phi}^*$ denotes the non dimensional form of $\phi$, $(u_1, u_2, u_3) = (u, v, w)$ are the velocity components in the $(x_1,x_2,x_3) = (x, y, z)$  direction, respectively, $t$ is time, $p$ is pressure, $\delta$ is the Kronecker delta, $f$ is the dynamically adjusted forcing term added to the momentum equation to ensure a constant flow rate, $c$ is the scalar concentration and $Sc$ is the Schmidt number.
The bulk Reynolds number is defined as $Re_b=U_b H/\nu$, 
where $H$ is the height of the channel and $U_b$ is the bulk velocity. The latter is defined by $U_b =\frac{1}{H}\int^H_0 \overline{\langle {u}\rangle} d{y}$, where $\overline{\,\cdot\,}$ and $\langle \cdot \rangle$ denotes averaging over time and the homogeneous horizontal ($x,z$) directions, respectively. 
The scalar concentration is normalised by 
\begin{equation}
  c^*=\frac { {c}-{c}_{b,0}}  {{c}_s-{c} _{b,0} },
\end{equation} 
 where ${c}_{b,0}$ is the initial concentration in the bulk and ${c}_{s}$ is the saturation concentration at the surface. 
 
The full set of governing equations was solved using the in-house KCFlo code \citep{kubrak:13}. 
The momentum equations were discretised using a fourth-order central scheme for diffusion combined with a fourth-order kinetic energy conserving scheme for convection. The Poisson equation for the pressure was solved using a conjugate gradient solver, with simple diagonal preconditioning. 
The scalar advection-diffusion equations were discretised using a fifth-order weighted essentially non-oscillatory (WENO) scheme \citep{Liu1994} for the convection and a fourth-order accurate central scheme for the diffusion.
As the scalar diffusivities are much smaller than the momentum diffusivity, a dual-mesh approach (using a finer mesh for the scalar computation than for the flow computation) was
used in order to accurately resolve the Batchelor scales in a computationally efficient manner. 

The solutions of both flow and scalar fields were advanced in time using the second-order accurate explicit Adams-Bashforth scheme.
Three scalar advection-diffusion equations with different Schmidt numbers were solved simultaneously, enabling a direct comparison of scalar transport processes driven by exactly the  same background turbulence.

The constitutive parts of our numerical code were validated in 
\cite{Wissink2004} for the flow solver and in \cite{kubrak:13} for the scalar advection-diffusion. In the latter reference, the numerical treatment of the scalar transport process was compared to analytical solutions for pure advection as well as pure diffusion, and the convergence rate of the method was rigorously established. Furthermore, the effectiveness and convergence of our dual-meshing approach was demonstrated in e.g. \cite{kubrak:13, herlina:14, wissink:16}.

%-------------------------------------------------------------------------
\subsection{Computational setup}
\label{sec:setup}
\begin{figure}
  \centering
  \includegraphics[scale=0.5, trim=0cm 1.5cm 0cm 0cm, clip]{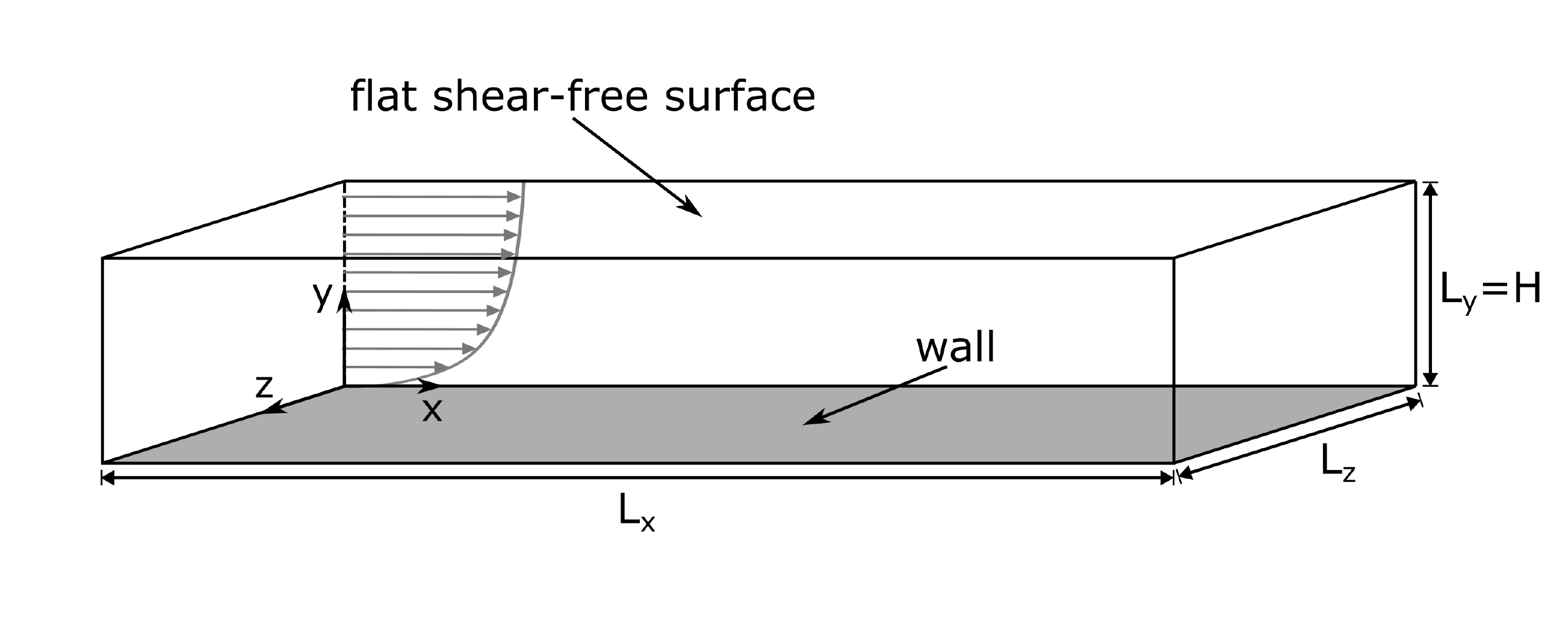}
  \caption{Schematic of the computational domain.}
  \label{fig:set}
\end{figure}

%-------------------------------------------------------------------------
As mentioned above, the present study focuses on low diffusivity mass transfer in a fully developed turbulent open channel flow. 
The size of the computational domain was $L_x\times L_y \times L_z$ in the streamwise ($x$), vertical ($y$) and spanwise ($z$) directions, respectively (cf.  figure~\ref{fig:set}).
Three different domain sizes were used, $3H\times H\times 3H$, $12H\times H\times 3H$ and $24H\times H\times 6H$. 
The total number of grid points employed in the simulations ranged from $4.72\cdot 10^6$ to $1.22\cdot 10^{10}$.
The Schmidt and bulk Reynolds numbers were varied from $Sc=4$ to $200$ and $Re_b=2875$ to $12000$, respectively. 
A list of the simulations performed is provided in Table~\ref{tab:mass}. 

\begin{table}
  {
  \begin{center}
    \rule{\textwidth}{.5pt}\vspace{2ex}
  \def~{\hphantom{0}}
  \small
  \begin{tabular}{c c c c c c c c c c c}
    Run & $Re_b$ & $Re_\tau$ & $Sc$ & $L_x\times L_y\times L_z$ & $N_x\times N_y\times N_z$ & $\psi_x\times \psi_y\times \psi_z$ & $\Delta t_f/t_b$ \\[3pt]
    G01 & $3000$ & $190$ & $8^R,16^R$ & $3H\times H \times 3H$ & $192\times 128 \times 192$ & $3\times 2 \times 3$ & 360\\
    G02 & $4000$ & $240$ & $8^R$ & $3H\times H \times 3H$ & $192\times 128 \times 192$ & $2\times 2 \times 2$ & $260$\\
    G03 & $5000$ & $290$ & $4^R,8^R$ & $3H\times H \times 3H$ & $192\times 128 \times 192$ & $3\times 3 \times 3$ & $380$\\
    G04 & $2875$ & $180$ & $7,64^R,100^R$ & $12H\times H \times 3H$ & $384\times 128 \times 192$ & $3\times 4 \times 3$ & $700$\\
    G05 & $4000$ & $240$ & $7,16^R,32^R$ & $12H\times H \times 3H$ & $384\times 128 \times 192$ & $2\times 2 \times 2$ & $250$\\
    G06 & $5000$ & $290$ & $7^R,16^R$ & $12H\times H \times 3H$ & $384\times 128 \times 192$ & $3\times 3 \times 3$ & $340$\\
    G07 & $3200$ & $200$ & $7,16^R,200^R$ & $24H\times H \times 6H$ & $1152\times 384 \times 1152$ & $6\times 2 \times 2$ & $120$\\
    G08 & $6400$ & $365$ & $7^R,16^R,100^R$ & $24H\times H \times 6H$ & $1152\times 384 \times 1152$ & $6\times 2 \times 2$ & $80$\\
    G09 & $12000$ & $630$ & $7^R,16^R,64^R$ & $24H\times H \times 6H$ & $1152\times 384 \times 1152$ & $6\times 2 \times 2$ & $100$\\
  \end{tabular}
  \end{center}
  \caption{
  Overview of the simulations. $Re_b$ is the bulk Reynolds number, $Re_\tau$ is the friction Reynolds number, $Sc$ is the Schmidt number, $H$ is the channel height,  $L_x \times L_y \times L_z$ denote the size of the domain in $x,\;y,\;z$ directions, respectively,  $N_x, N_y, N_z$ are the number of grid points  of the base mesh, while $\psi_x, \psi_y, \psi_z$ denote the refinement factors used for the finer scalar mesh (of the $Sc$ cases marked with superscript $R$) and $\Delta t_f /t_b$ is the time-window used for the flow statistics, where $t_b=H/U_b$ is the bulk time unit. Note that the time-window of the scalar statistics differ from $\Delta t_f /t_b$, as will be explained in section ~\ref{sec:scalar}.\\[-.5ex]
  \noindent\hspace{-30pt}\rule{\textwidth}{.5pt}}
  \label{tab:mass}
  }
 \end{table}

In all simulations, the surface was assumed to be flat, thereby neglecting any possible influence of waves on mass transfer. 
For both flow and scalar fields, periodic boundary conditions were used in the streamwise and spanwise directions. 
In the vertical direction, a free-slip boundary condition at the free-surface ($y/H=1$) and a no-slip boundary condition at the wall ($y/H=0$) of the channel were applied for the velocity field. For the scalars, at the surface a Dirichlet boundary condition with  $c=c_s$, modeling the presence of the atmosphere, was employed, while at the wall a zero flux boundary condition ($\partial c/ \partial y = 0$) was used. 
To speed-up the formation of a turbulent concentration boundary layer, the concentration field was initialised by 
\begin{equation}
   c^*(1-y/H)=\mbox{erfc}\left(\frac{1-y/H}{\sqrt{{4t_d^*}/{(Re_bSc)}}}\right),
\label{eq:initialscalar}
\end{equation}
where $t_d^*$ is the non-dimensional diffusion time, which was set to $10$ for the smallest domain considered and $12$ for the others. 
 
The spatial discretisation was performed on a non-uniform and staggered Cartesian mesh. 
The mesh was uniform in the homogeneous ($x$, $z$) directions and stretched in the vertical ($y$) direction with refinement near the free-surface, located at $y(N_y)=H$, and the wall, located at $y(0)=0$. 
The node distribution is given by
\begin{eqnarray}
 y(k) = \left[1-\frac{\tanh(y_\phi)}{\tanh(y_1)}\right]\frac{y(0)}{2}
                             +  \left[1+\frac{\tanh(y_\phi)}{\tanh(y_1)}\right]\frac{y(N_y)}{2}
\label{eqn:stretching1}
\end{eqnarray}
for $k=1,...,N_y-1$, with 
\begin{eqnarray}
 y_1&=&\psi_m/2, 
 \\
 y_{\phi}&=&\psi_m \left[\frac{k}{N_y}-0.5\right] 
\end{eqnarray}
where $N_y$ is the number of nodes in the $y$-direction.
The stretching is controlled by the parameter $\psi_m = \frac{k}{N_y} \psi_t + \left[1-\frac{k}{N_y}\right]\psi_b$, where $(\psi_t, \psi_b)$ was set to $(2, 2)$ in the smallest domain, $(3, 2)$ in the mid-sized domain and $(2.3, 3)$ in the largest domain.
The adequacy of the base mesh resolution for all simulations was confirmed by studying   one-dimensional energy spectra of the velocity,  as discussed in section~\ref{sec:flow}.

As mentioned above, a dual-mesh strategy was employed to accurately resolve the evolution of the high $Sc$ scalar fields.
The refined grid resolution was chosen such that (i) the vertical grid size $\Delta y$ at the surface was less than $0.15L_B$, where $L_B=\eta/\sqrt{Sc}$ is the Batchelor scale and $\eta$ is the Kolmogorov scale, and (ii) the geometric mean of the grid cells $\Delta=\sqrt[3]{\Delta x\Delta y \Delta z}$ was $\Delta<\pi\,L_B$ in the upper part of the domain ($y/H \le 0.55$). 
These two conditions fulfil the criterion proposed by \cite{Groetzbach1983}, which ensures that the scalar mass transfer in all simulations is adequately resolved.

%-------------------------------------------------------------------------
\subsection{Domain size} 
%-------------------------------------------------------------------------
\label{sec:domain_size}
\begin{figure}
  \sidesubfloat[]
  {\includegraphics[scale=.14]{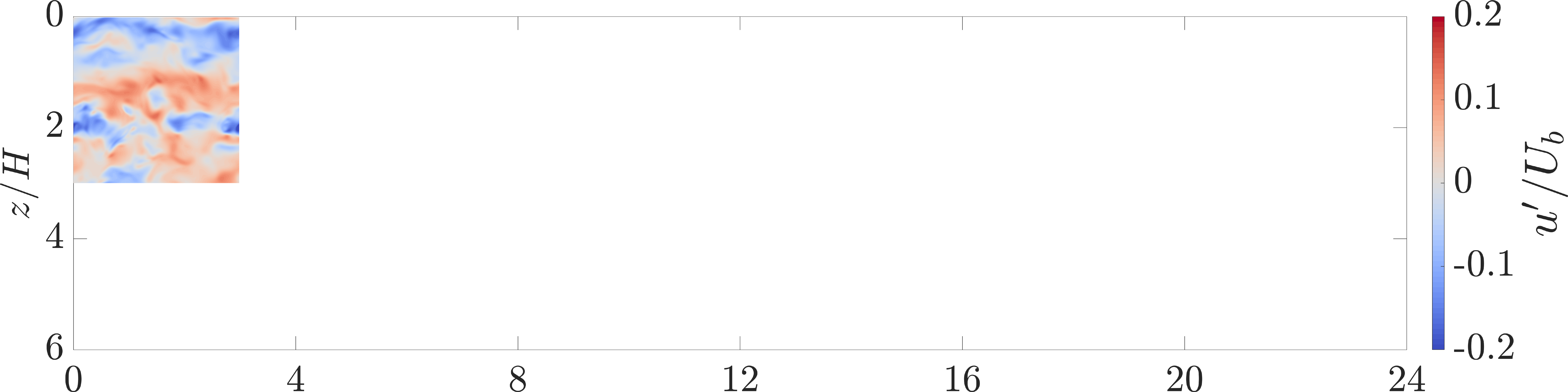}
  \label{fig:snap1}}
  \sidesubfloat[]
  {\includegraphics{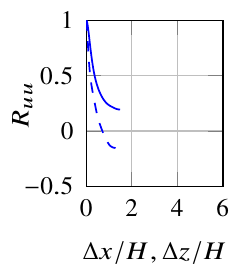}    
  \label{fig:corrall1}} \\ 
  \sidesubfloat[]
  {\includegraphics[scale=.14]{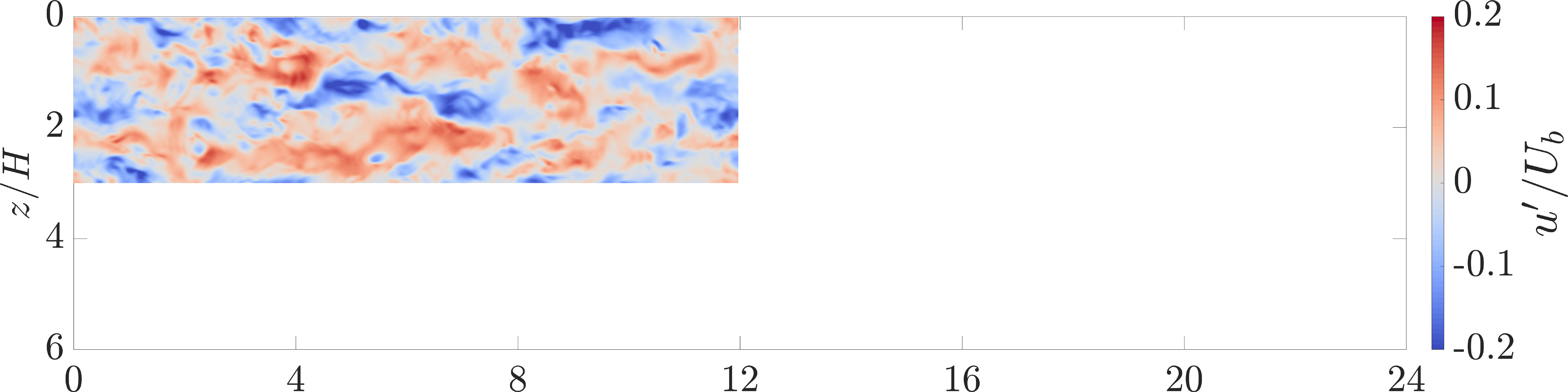}
  \label{fig:snap2}} 
  \sidesubfloat[]
   {\includegraphics{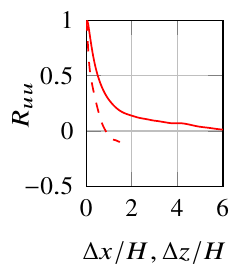} 
  \label{fig:corrall2}} \\ 
  \sidesubfloat[]
  {\includegraphics[scale=.14]{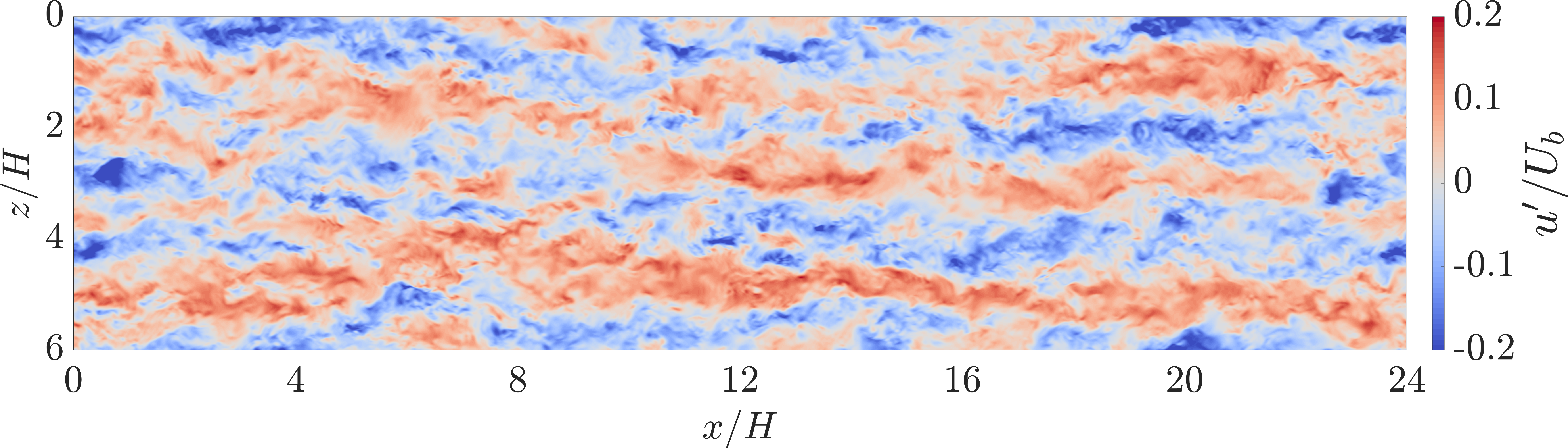}
  \label{fig:snap3}} 
  \sidesubfloat[]
   {\includegraphics{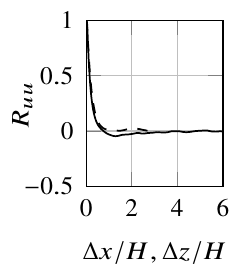} 
  \label{fig:corrall3}}
 \caption{
 Instantaneous contour maps of $u'/U_b$ (left panes) and time-averaged two-point correlations of the streamwise velocity (right panes) in the plane $y/H=0.6$ for \protect\subref{fig:snap1}-\protect\subref{fig:corrall2}  $Re_\tau=290$ and  \protect\subref{fig:snap3}, \protect\subref{fig:corrall3}  $Re_\tau=630$.  The lines in \protect\subref{fig:corrall1}, \protect\subref{fig:corrall2} and \protect\subref{fig:corrall3} represent $R_{uu}$ in streamwise (solid lines) and spanwise (dashed lines) directions. 
  \label{fig:corrplusfig}
  }
\end{figure}

Before proceeding with the discussion of the results, a brief overview of the effect of the domain size on the flow structures in open channel flow is presented. 
A more detailed investigation can be found in e.g. \cite{Wang2020, Bauer2020}, while  
similar investigations for closed channel flow were carried out by e.g. 
\cite{HWang2010,Lozano2014,Feldmann2018,Abe2018}.

One of the main findings reported in the investigations mentioned above is the appearance of very large coherent structures for large $Re_\tau$. To assess the suitability of the domain sizes used in the present simulations for capturing such large coherent structures, typical snapshots of the streamwise velocity fluctuations $u^\prime$ at 
$y/H=0.6$ for G03, G06, G09 are shown in figures~\ref{fig:snap1}, \ref{fig:snap2} and \ref{fig:snap3}, respectively.
As can be seen in table \ref{tab:mass}, these cases have the highest $Re_\tau$ for each of the three domain sizes considered.
In all figures, high and low velocity elongated streaky structures, that are characteristic of open-channel flow, can be seen. The snapshots and corresponding two-point correlations 
\begin{eqnarray}
R_{uu}^x(\Delta x,y)=\overline{ \left ( \frac{\langle u^\prime(x,y,z,t) u^\prime(x+\Delta x,y,z,t) \rangle} {\langle u^\prime(x,y,z,t)^2 \rangle}  \right) }
\\
R_{uu}^z(\Delta z,y)=\overline{ \left ( \frac{\langle u^\prime(x,y,z,t) u^\prime(x,y,z+\Delta z,t) \rangle} {\langle u^\prime(x,y,z,t)^2 \rangle} \right) }
\end{eqnarray}
at $y/H=0.6$
(figures \ref{fig:corrall1}, \ref{fig:corrall2} and \ref{fig:corrall3}) indicate that the streamwise extent of these coherent structures was captured quite well in the largest domain and marginally in the mid-sized domain, but not in the smallest domain. 

For both $Re_\tau=365$ and $630$, the proper de-correlation of $u^\prime$ in the largest domain  (shown for $Re_\tau=630$ and $y/H=0.6$ in figure \ref{fig:corrall3}) was achieved in the spanwise direction for all $y/H$ and in the streamwise direction for $y/H \leq 0.7$. 
For $y/H>0.7$ the minimum value for the streamwise 
$R_{uu}$ was always smaller than $0.05$, indicating a more marginal de-correlation. In contrast, for the lowest $Re_\tau=200$, a de-correlation of $u^\prime$ in both horizontal directions was obtained for the entire depth.  
In the mid-sized simulations, the streamwise de-correlation was  marginal for all cases, while in the smallest domain no de-correlation was achieved. 
As expected, a domain size of $3H \times H \times 3H$ is too small to fully capture turbulent open-channel flow, even for $Re_\tau$ as low as $190$. 
Nevertheless, it will be shown in section \ref{sec:mass} that for $Re_\tau = 240$ and $290$ the interfacial mass transfer was found to be largely independent of the domain size.

Furthermore, in figure \ref{fig:snap3} a coherent structure of length $\approx 20H$ can be observed. 
In the literature, structures of length $\gtrsim 10H$ are usually referred to as very large scale motions (VLSM), while the term large scale motions (LSM) is typically used for structures with a length of $\approx 1-3H$. 
In open channel flow, the onset of the appearance of VLSM is still uncertain, i.e. to date such motions were confirmed to appear for $Re_\tau\geq  400$ \citep {Bauer2020},  $Re_\tau\ge  550$  \citep {Wang2019} and experimentally at  $Re_\tau\geq  700$  \citep{Peruzzi2020}.
The very large coherent structure seen in  figure \ref{fig:snap3} is an example of such a very large scale motion that is found at $Re_\tau=630$. Later it will be shown that in the present simulations VLSM could already be detected at $Re_\tau = 365$. Further confirmation of the appearance of VLSM based on e.g. pre-multiplied spectra and velocity fluctuations analyses, is presented in section \ref{sec:flow}. 

%-------------------------------------------------------------------------
\section{Flow characteristics}
%-------------------------------------------------------------------------
\subsection{Flow statistics}
%-------------------------------------------------------------------------
\label{sec:flow}
All simulations were started from fully developed turbulent flow fields.
The shown statistics, if not stated otherwise, were obtained by averaging in the homogeneous ($x$, $z$) directions and in time (cf. table~\ref{tab:mass}). 

%--------------------------------
\setlength{\labelsep}{-0.3cm}  
\begin{figure}
   \sidesubfloat[]{
   \includegraphics{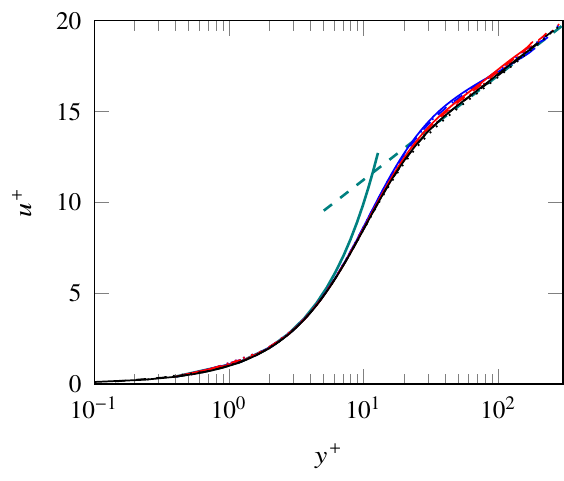}
   \label{fig:uplus}} \quad
   \sidesubfloat[]{
   \includegraphics{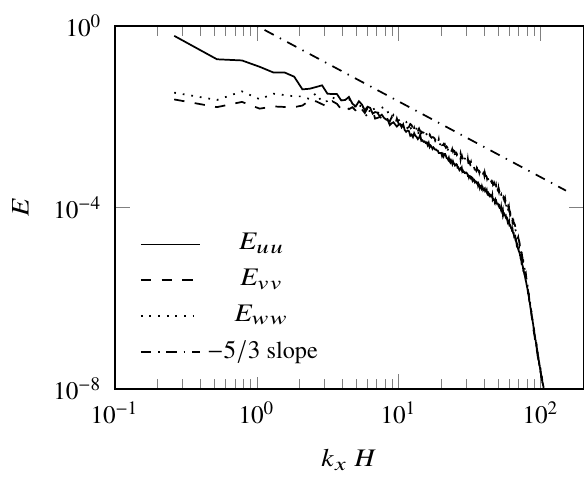}
   \label{fig:1dspectra}}    
      \caption{
      \protect\subref{fig:uplus} 
      Near-wall mean streamwise velocity profiles~: 
      \protect\myline{blue}, G01;     
      \protect\myline{blue,dashed}, G02;     
      \protect\myline{blue,dotted}, G03;     
      \protect\myline{red}, G04;     
      \protect\myline{red,dashed}, G05;     
      \protect\myline{red,dotted}, G06;     
      \protect\myline{black}, G07;     
      \protect\myline{black,dashed}, G08;     
      \protect\myline{black,dotted}, G09;
      \protect\myline{teal}, $u^+=y^+$;  
      \protect\myline{teal,dashed}, the logarithmic law  with $\kappa=0.40$ and $B=5$.       
      \protect\subref{fig:1dspectra} One-dimensional energy spectra in streamwise direction from simulation G09 at $y/H=0.7$.
      }
      \label{fig:flow_stat}
\end{figure}
%---------------------------------
Figure~\ref{fig:uplus} shows that the mean streamwise velocity profiles in all  simulations agree well with the law of the wall, including a logarithmic region $u^+=\frac{1}{\kappa}log(y^+)+B$, with standard values for the von Karman constant ($\kappa=0.40$) and for the intercept ($B=5$). The one-dimensional streamwise energy spectrum of the velocity components for the simulation with the largest Reynolds number (G09) is shown in figure~\ref{fig:1dspectra}.  As also observed for all other simulations, the existence of an inertial subrange, indicated by the $k^{-5/3}$ power law, can be clearly seen. 
Furthermore, no energy pile-up at high wavenumbers was observed, demonstrating that the smallest scales of motion were well resolved in all simulations.
In addition, it was found that  $Re_\tau=0.166Re_b^{0.88}$, which is in agreement with the results for closed channel flow as shown by \cite{Pope2000} and \cite{Lee2015}. 

%----------------------------
\begin{figure}
   \setlength{\labelsep}{-0.01cm}
    \sidesubfloat[][$Re_\tau=200$.]
      {\includegraphics[height=.2\textheight]{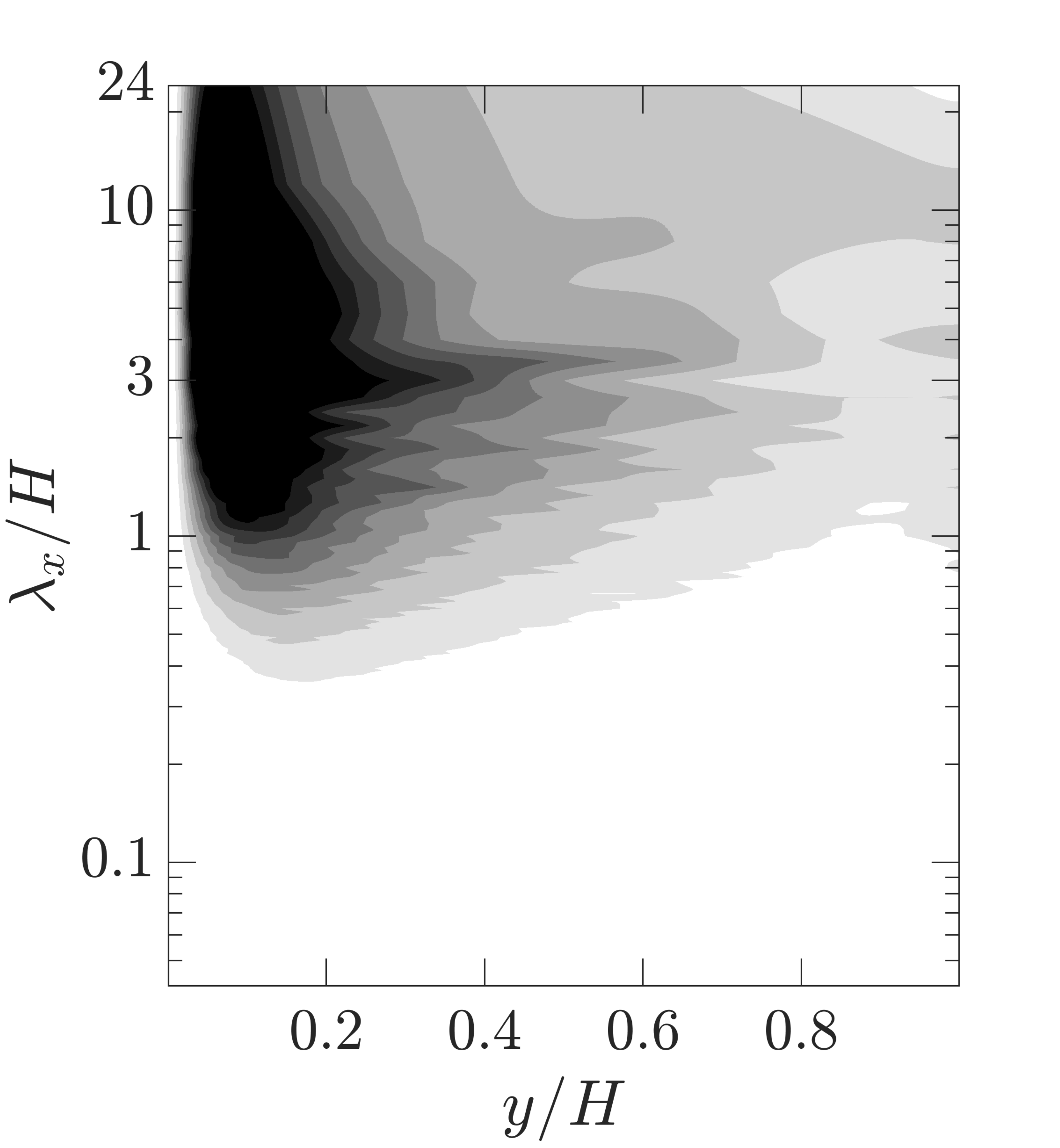}\label{fig:EuuxG07}
      }
   \sidesubfloat[][$Re_\tau=365$.]
      {\includegraphics[height=.2\textheight]{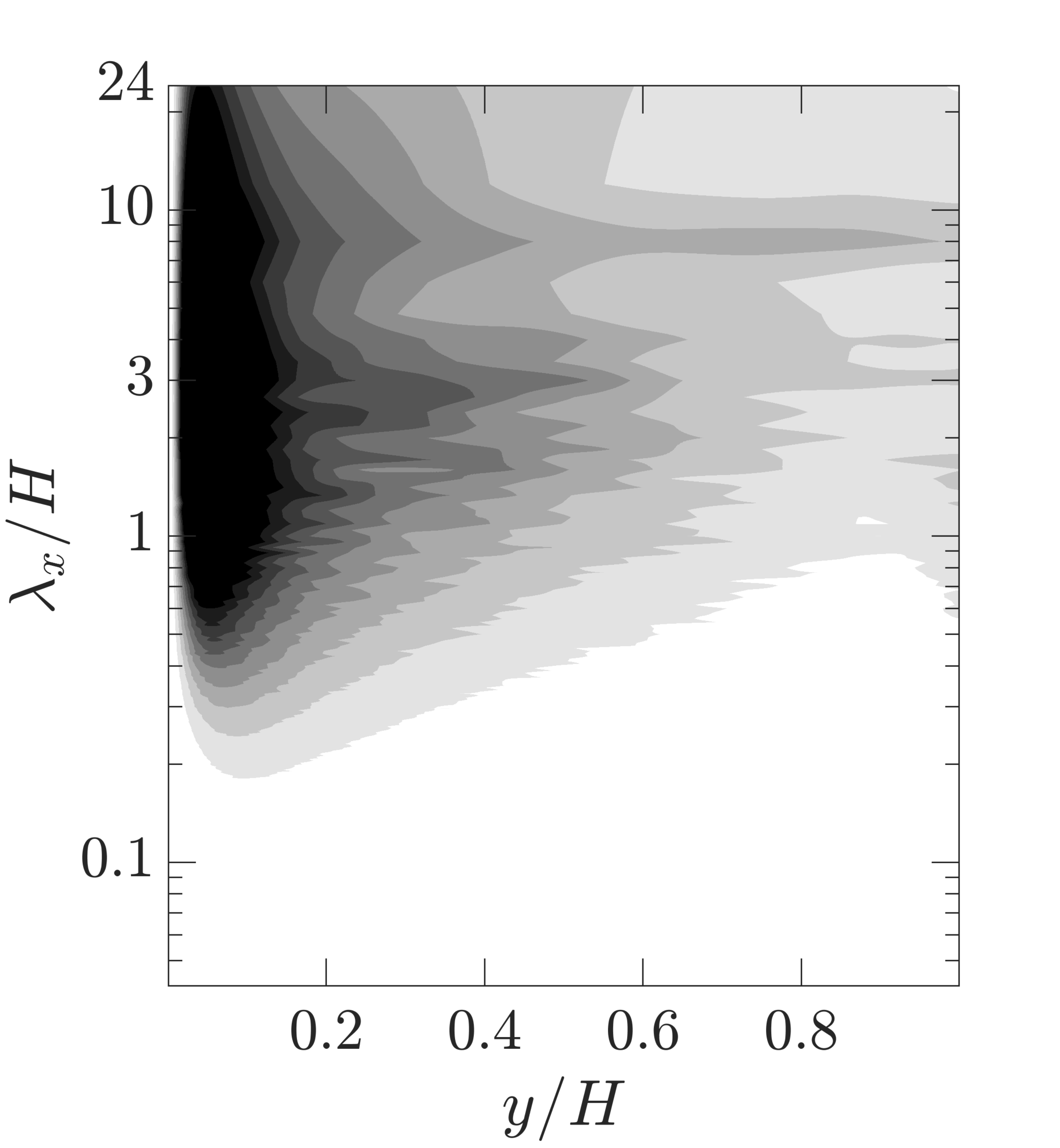}\label{fig:EuuxG08}
      }
      \sidesubfloat[][$Re_\tau=630$.]
      {\includegraphics[height=.2\textheight]{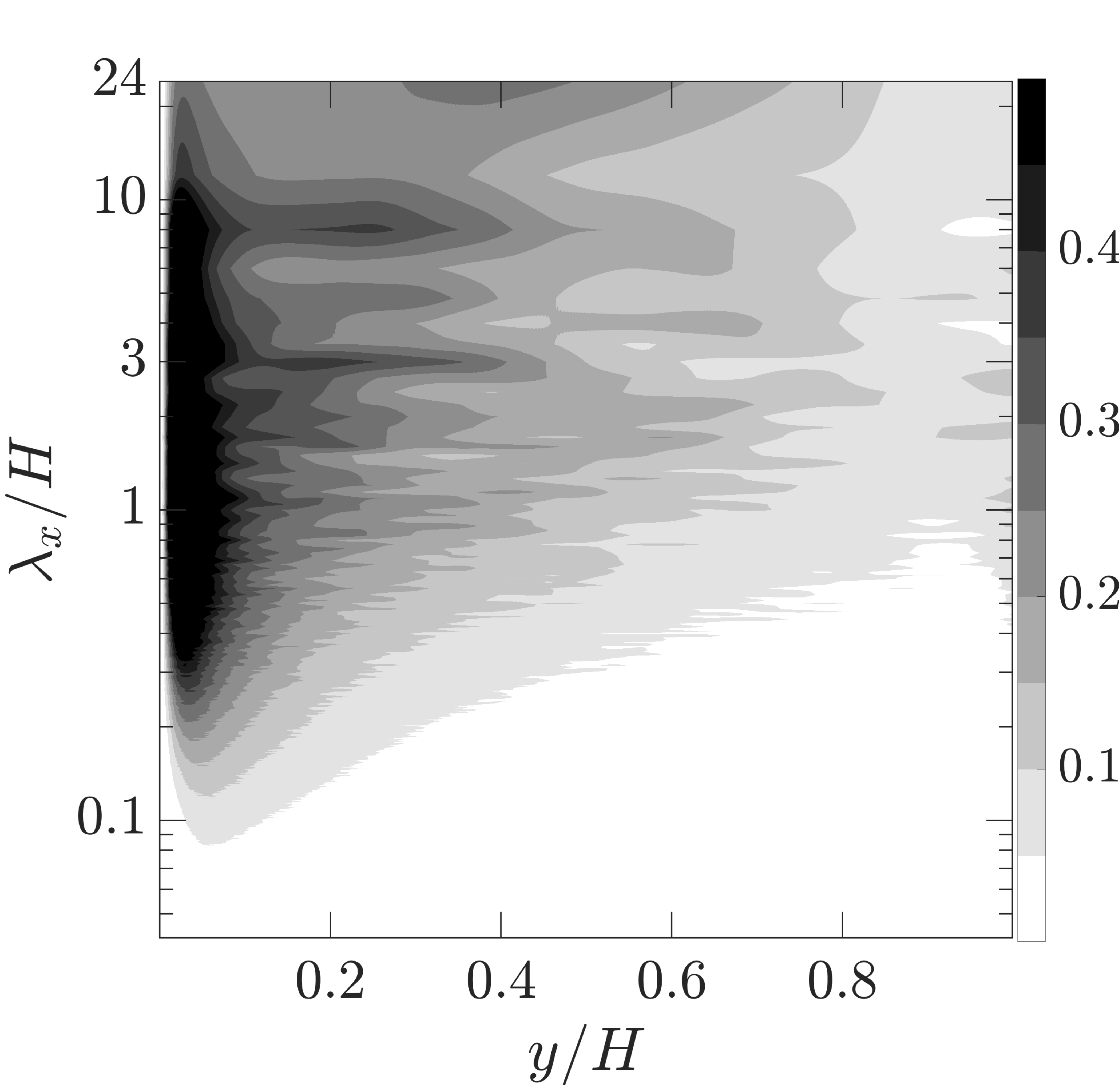}\label{fig:EuuxG09}
      }\\
     \sidesubfloat[][$Re_\tau=200$.]
      {\includegraphics[height=.152\textheight]{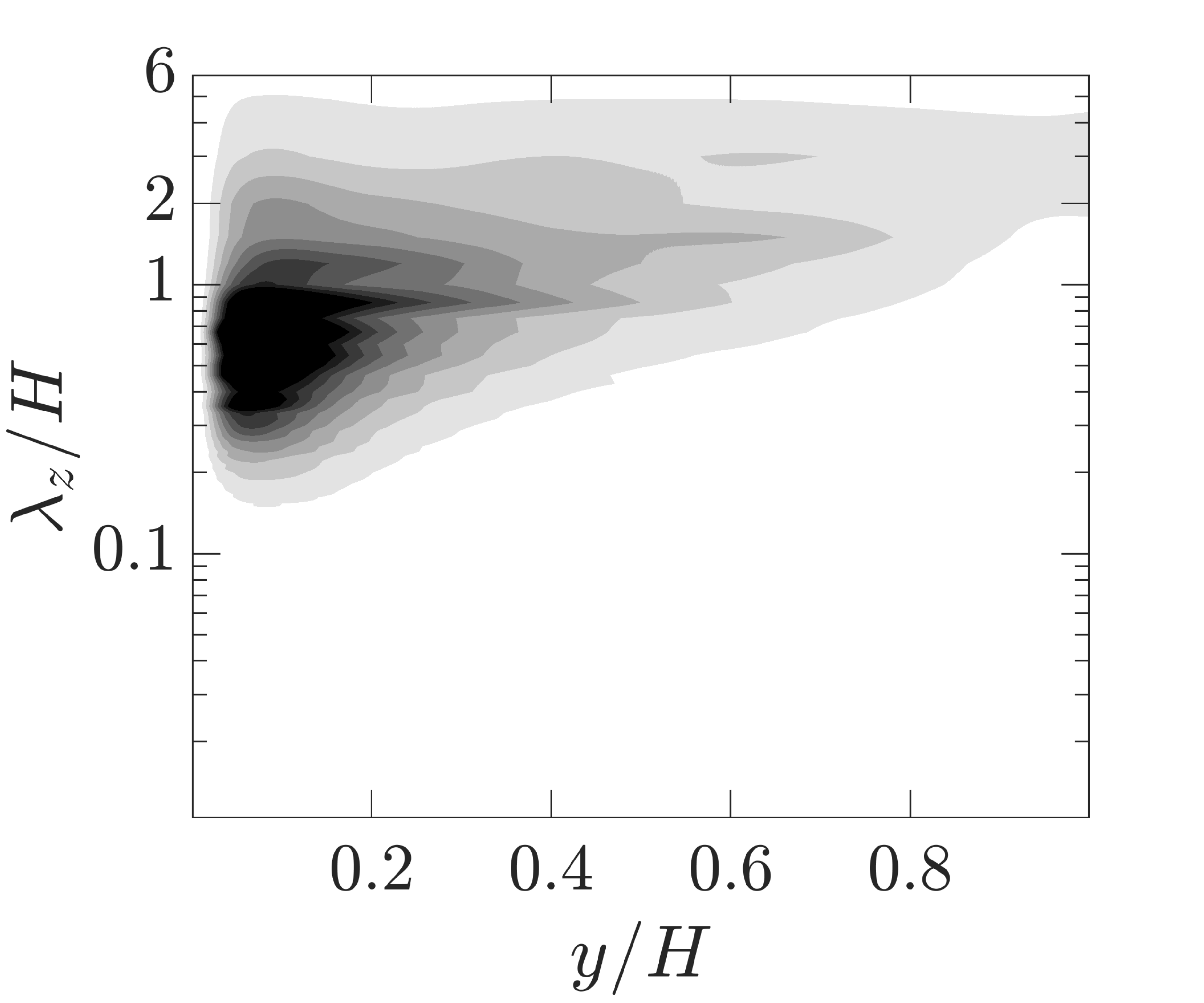}\label{fig:EuuyG07}
      }
   \sidesubfloat[][$Re_\tau=365$.]
      {\includegraphics[height=.152\textheight]{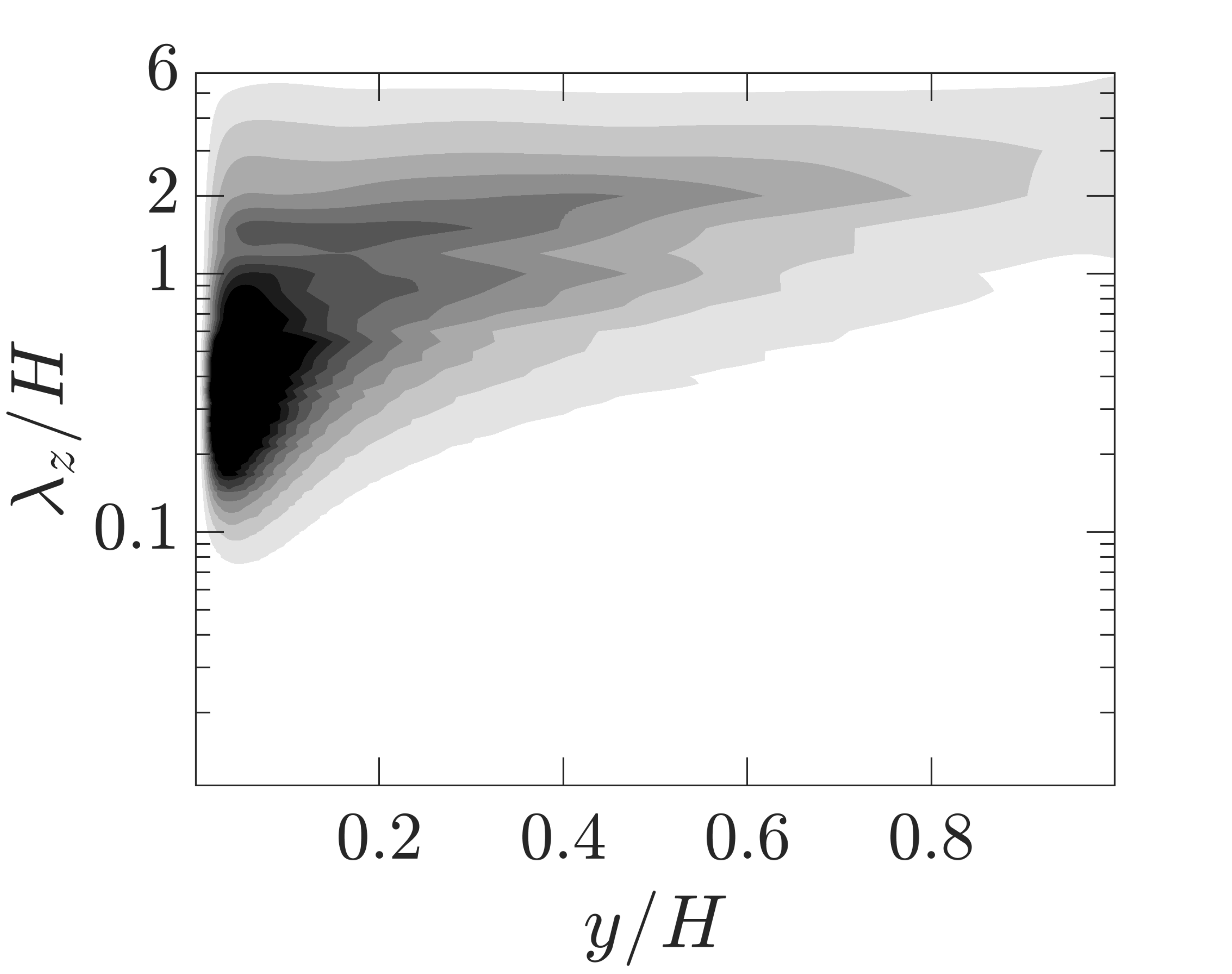}\label{fig:EuuyG08}
      }
      \sidesubfloat[][$Re_\tau=630$.]
      {\includegraphics[height=.152\textheight]{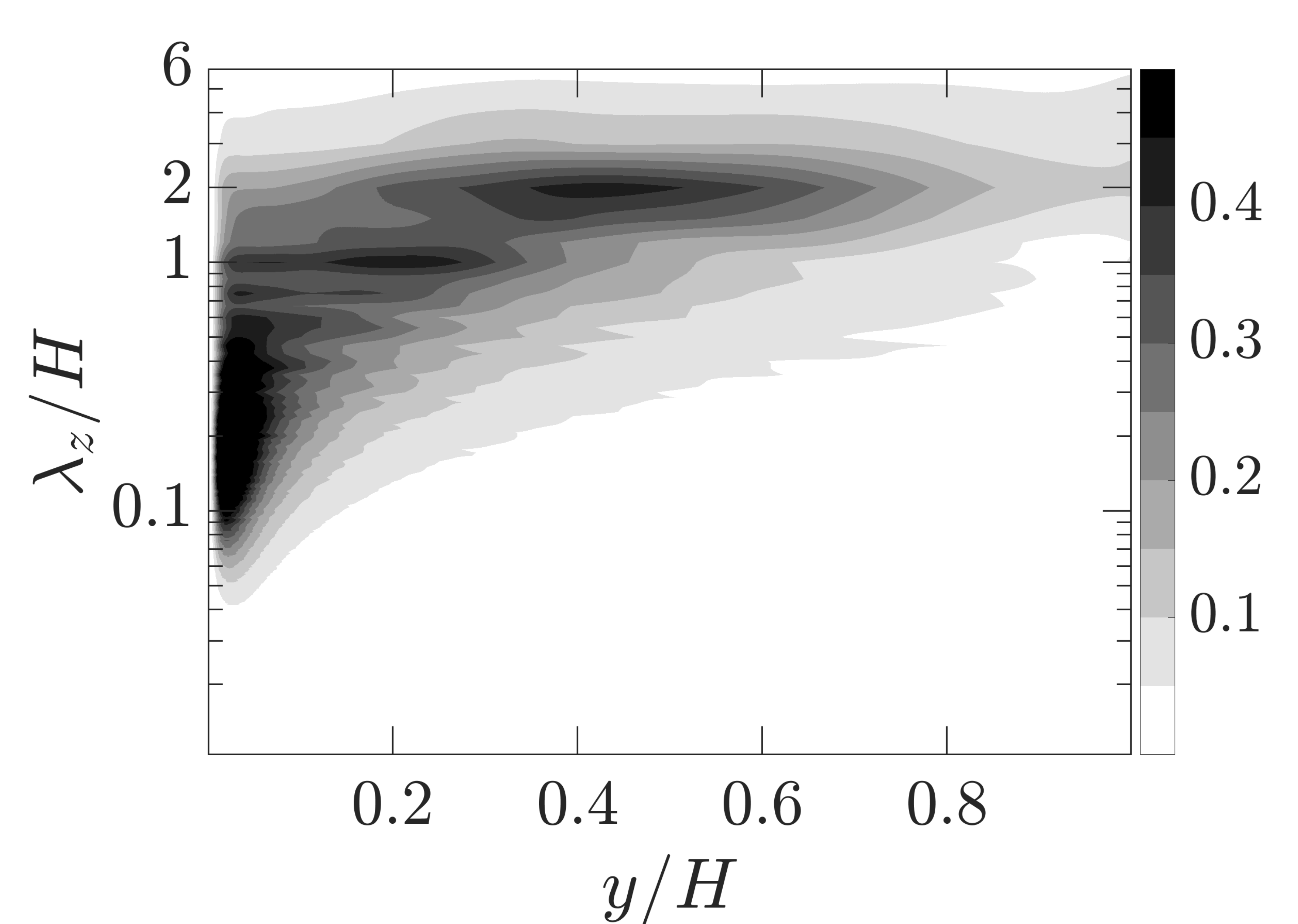}\label{fig:EuuyG09}
      }
   \caption{
   Contour maps of normalised pre-multiplied longitudinal velocity fluctuation spectra 
   in \protect\subref{fig:EuuxG07}-\protect\subref{fig:EuuxG09} streamwise ($E^*_{u'u'}=k_x\,E_{u'u'}/(k_x\,E_{u'u'})_{max})$) and \protect\subref{fig:EuuyG07}-\protect\subref{fig:EuuyG09} 
   spanwise ($E^*_{w'w'}=k_z\,E_{w'w'}/(k_z\,E_{w'w'})_{max})$) directions as a function of 
   wavelength and distance from the wall ($y/H$). The Reynolds numbers for the shown cases are 
   \protect\subref{fig:EuuxG07}, \protect\subref{fig:EuuyG07} $Re_\tau=200$, \protect\subref{fig:EuuxG08}, \protect\subref{fig:EuuyG08} $Re_\tau=365$ and \protect\subref{fig:EuuxG09}, \protect\subref{fig:EuuyG09} $Re_\tau=630$. 
   }
   \label{fig:Euu}
\end{figure}  
%------------------------------------
Figure~\ref{fig:Euu} shows contours of the longitudinal velocity fluctuation spectra in the streamwise ($x$) and the spanwise ($z$) direction from the simulations performed in the $24H \times H \times 6H$ domain (G07-G09). The energy spectra were premultiplied by the wavenumber and are shown as a function of the  wavelength and distance from the wall. 
As expected, for increasing Reynolds number the amount of energy at smaller wavelengths was found to increase, especially near the wall of the channel. 
In the streamwise spectra  (figures~\ref{fig:EuuxG07}, \ref{fig:EuuxG08}, \ref{fig:EuuxG09}), an energy peak at a wavelength of $\lambda_x/H \approx 3 $ was identified in all three cases.  This peak is associated with LSM. 
It should be noted that in the spectra, the location of the peaks with higher wavelengths becomes less accurate due to the limited size of the computational domain $L_x=24H$, $L_z=6H$.
Even though the location may not be entirely correct, the energy peaks observed at $\lambda_x \gtrsim 10H$, which extend over  virtually the whole channel height, indicate the presence of VLSM for $Re_\tau \ge 365$ (G08, G09).
When examining the spanwise spectra (figures~\ref{fig:EuuyG07}, \ref{fig:EuuyG08}, \ref{fig:EuuyG09}),
energy peaks at $\lambda_z/H \approx 1$, which relate to LSM, were detected in all three cases. 
High energy values at $\lambda_z/H \ge 2$ (typical for VLSM) were observed for $Re_\tau=365$ and $Re_\tau=630$ when $y/H\ge 0.3$ and $0.1$, respectively.  

%------------------------------
\begin{figure}
  \setlength{\labelsep}{0cm}
  \centering
  \sidesubfloat[]{
    \includegraphics{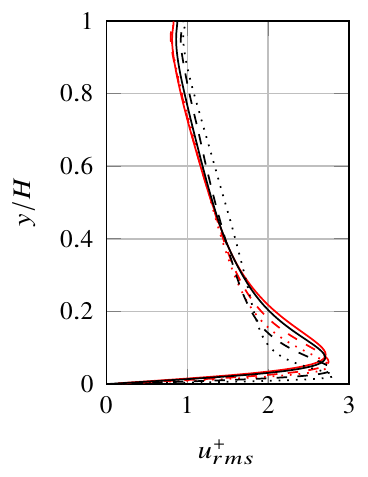}
    \label{fig:rms1}}\quad
  \sidesubfloat[]{
    \includegraphics{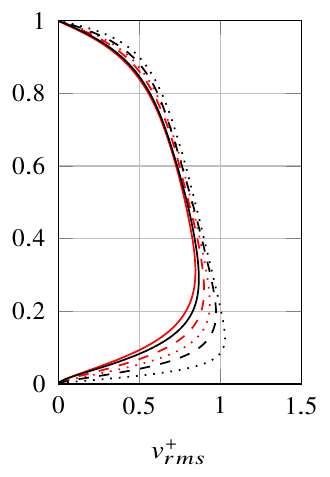}  
    \label{fig:rms2}}\quad
      \sidesubfloat[]{
    \includegraphics{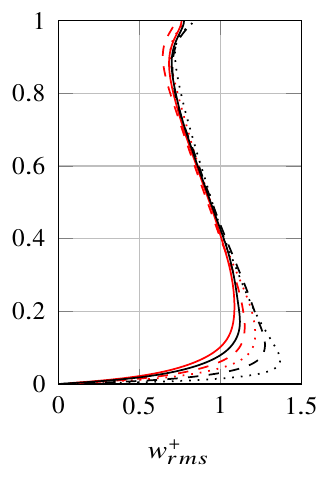}  
    \label{fig:rms3}}
      \caption{Profiles of root mean square velocity fluctuations normalised by the friction velocity $u_\tau$ as a function of $y/H$. The lines represent 
      \protect\myline{red,solid}, G04; 
      \protect\myline{red,dashed}, G05; 
      \protect\myline{red,dotted}, G06; 
      \protect\myline{black,solid}, G07; 
      \protect\myline{black,dashed}, G08; 
      \protect\myline{black,dotted}, G09.}
  \label{fig:rms}
\end{figure}
%------------------------------

The root mean square (r.m.s.) of the velocity fluctuations for the simulations performed using the mid- and large sized domains are shown in figure~\ref{fig:rms}. 
As expected, a decrease in the boundary layer thickness at the wall with increasing Reynolds number was observed. 
The increase in the peak of the streamwise component $u_{rms}$ (figure~\ref{fig:rms1}) from $2.7$ for $Re_\tau=200$ to $2.8$ for $Re_\tau=630$ is in  good agreement with the values reported in \cite{Kim1987,Pope2000,Wang2019}. 
For $y/H \ge  0.3$,  the $u_{rms}^+$ profiles for $Re_\tau < 365$ were found to nearly collapse on one curve, while for $Re_\tau \ge 365$ $u_{rms}^+$ tends to grow with $Re_\tau$. This increase in $u_{rms}^+$
tends to be associated to the presence of VLSM \citep{Kim1999,DelAlamo2004,Hoyas2006}. 

Figures~\ref{fig:rms2} and \ref{fig:rms3} show that for $y/H \ge 0.3$, similar $v_{rms}^+$ profiles and similar $w_{rms}^+$ profiles were obtained for all Reynolds numbers.
At the surface, vertical fluctuations are damped and the turbulent kinetic energy is redistributed in the horizontal directions, which explains the increase observed near the surface in the $u_{rms}^+$ and $w_{rms}^+$ profiles. 
The thickness of the surface influenced layer $L_{s}$ was determined by the distance between the surface and the location, $y_\infty$, where
$I(y)=
\overline{
\left( {\langle uu\rangle}+{\langle vv \rangle}+{\langle ww\rangle}  \right) / 
\left( {\langle uu\rangle}+{\langle ww\rangle} \right) }
$
is maximum ($L_s = (H-y_\infty)$). $I$  can be seen as a measure of the Reynolds stress anisotropy, where the lower limit  $I=1$ corresponds to 2D flow, while isotropic flow would result in $I=3/2$. In open channel flow, the former is achieved at the free-surface. With increasing distance from the surface ($H-y > 0$), the flow becomes less anisotropic and hence the magnitude of $I$ increases until it reaches a maximum at the edge of the surface influenced layer ($y=y_\infty$). Here, in all simulations, $y_\infty$ was found to be located somewhere between $0.6H$ and $0.75H$  (cf. table \ref{tab:ReT}), such that $0.25H\leq L_s \leq0.4H$. When comparing simulations of the same domain size, the surface influenced layer $L_s$ was found to decrease with increasing $Re_\tau$.

%---------------------------------------
\begin{figure}
\setlength{\labelsep}{0cm}
\centering
  \sidesubfloat[]{\includegraphics{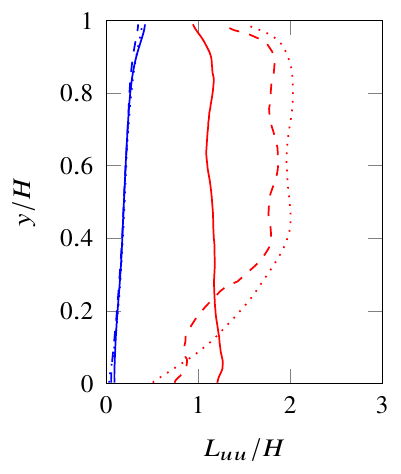}  \label{fig:L_uu}}
  \sidesubfloat[]{\includegraphics{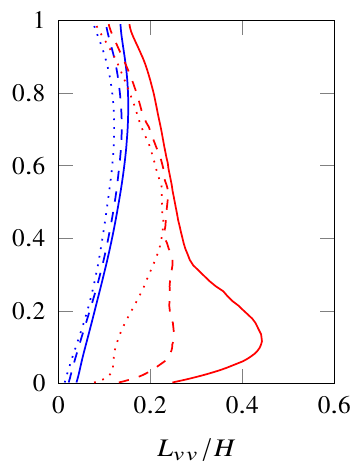}  \label{fig:L_vv}}
  \sidesubfloat[]{\includegraphics{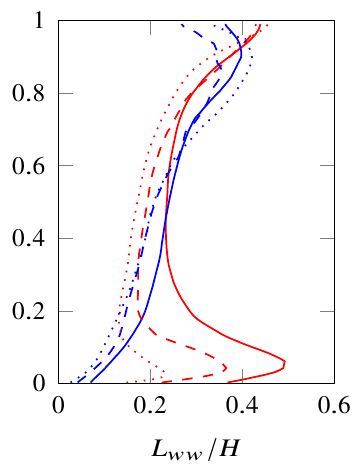}  \label{fig:L_ww}}
  \caption{Integral length scales of the velocity components \protect\subref{fig:L_uu} $u$, 
  \protect\subref{fig:L_vv} $v$ and \protect\subref{fig:L_ww} $w$  in the $x$-direction (red) and $z$-direction (blue) as a function of $y/H$.  The data are from the $24H \times H \times 6H$ domain simulations with $Re_\tau=200$ (solid lines),  $Re_\tau=365$ (dashed lines) and $Re_\tau=630$ (dotted lines). }
  \label{fig:int_lengths}
\end{figure}
%---------------------------------------
Figure~\ref{fig:int_lengths} shows the integral length scales for the velocity components in the homogeneous directions as a function of $y/H$ for the largest domain size. 
For $y/H > 0.4$, a significant increase in the integral length scale in the $x-$direction ($L_{uu}^x$ ) by about 
$50\%$ was observed when increasing the Reynolds number from $Re_\tau=200$ to $365$. 
Further increasing $Re_\tau$ to $630$ resulted in a further increase in $L_{uu}^x$ by about $10\%$ (figure \ref{fig:L_uu}). 
This significant growth in $L_{uu}^x$ for $Re_\tau=365$ and $630$ corresponds to the presence of VLSM. 
For $y/H\leq0.05$, all integral length scales in the $x$-direction ($L_{uu}^x$, $L_{vv}^x$ and $L_{ww}^x$) can be seen to decrease with increasing Reynolds number.   
This trend persists for $L_{vv}^x$ for (almost) every $y/H$, and for $L_{ww}^x$ until $y/H\approx 0.8$. 
Compared to the $x-$direction, with the possible exception of $L_{ww}^z$, the integral length scales in the $z-$direction do not show any significant Reynolds number effect. 
 
%-------------------------------------------------------------------------
\subsection{Flow structures}
%----------------------------------------------------------------------------
\label{sec:VLSM}

%----------
\setlength{\labelsep}{0cm}
\begin{figure}
  \includegraphics[width=.99\textwidth,]{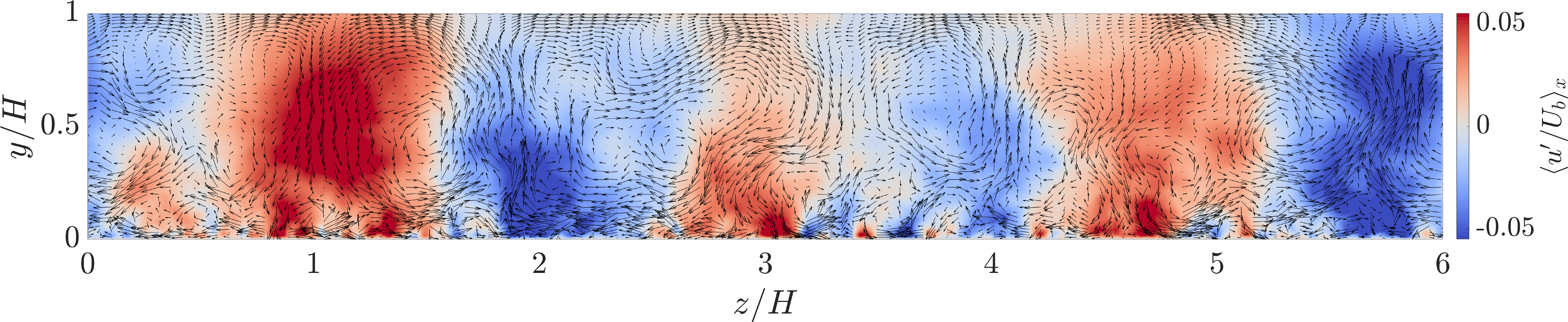}
  \caption{Typical contour plot of the streamwise averaged $u-$fluctuation $\langle u^\prime /U_b \rangle_x$ from simulation G09 (at $t/t_b=52$). The black arrows represent the streamwise averaged velocity components in the cross-plane, ($\langle v/U_b \rangle_x, \langle w/U_b \rangle_x$).} 
  \label{fig:avgx_flow}
\end{figure}
%----------

Figure~\ref{fig:avgx_flow} shows contours of the streamwise-averaged $u$-fluctuation ($\langle u^\prime /U_b \rangle_x$), together with  streamwise-averaged velocity vectors of the two components in the plane. Large areas with high and low speed streamwise flow can be observed, which extend almost from the wall to the free-surface of the channel. Downward moving flow is typically present in the high streamwise velocity areas, while in the low-speed areas the flow tends to move upwards. 
It was observed in figure~\ref{fig:snap1} that these high and low speed areas extend over a significant streamwise portion of the channel, and are related to VLSM. 

\begin{figure}
\centering
  \sidesubfloat[][$y/H\ge 0.5$, $Q/Q_{rms}=1$.]
      {\includegraphics[scale=0.2]{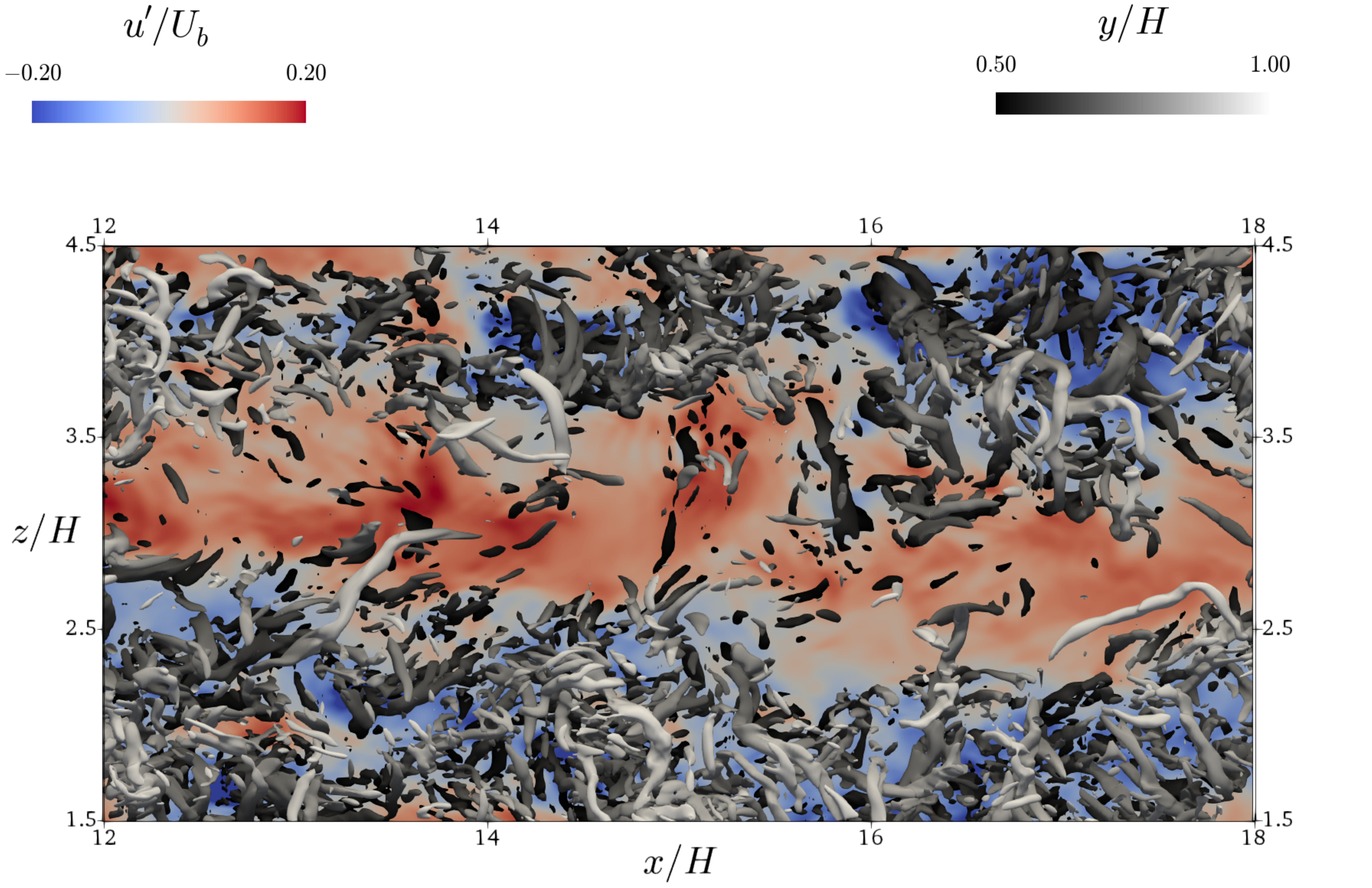}
      \label{fig:Q1}}\\
  \sidesubfloat[][$y/H\ge 0.8$, $Q/Q_{rms}=0.1$.]
      {\includegraphics[scale=0.2]{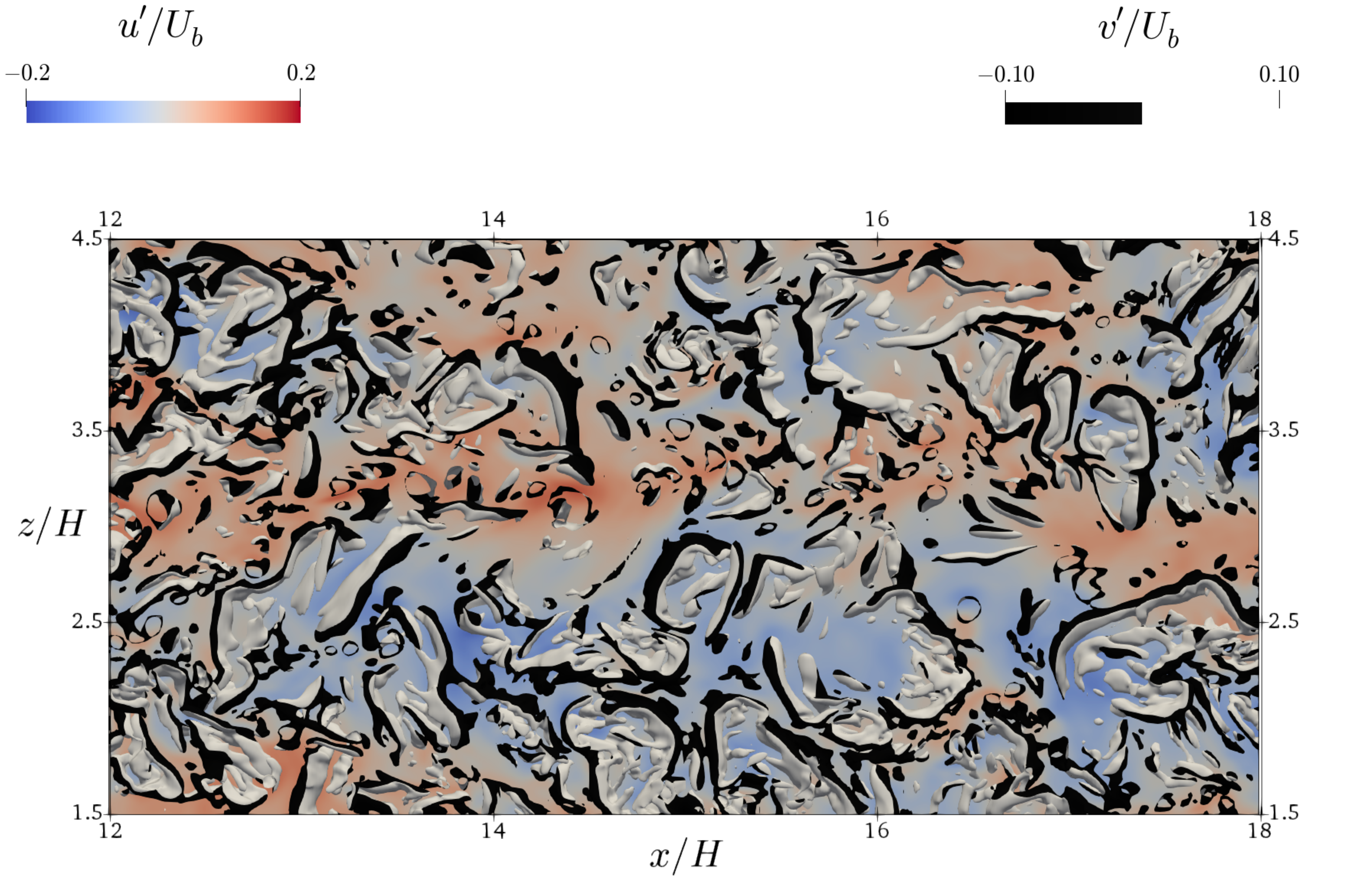}
      \label{fig:Q2}}
  \caption{Typical vortical structures from simulation G09 (at $t/t_b\,=\, 52$), visualised by iso-surfaces that correspond to a value of \protect\subref{fig:Q1} $q=1$  and 
   \protect\subref{fig:Q2} $q=0.1$, see equation (\ref{eq:Qnorm}). In 
   \protect\subref{fig:Q1} the structures are coloured by $y/H$, while in 
   \protect\subref{fig:Q2} they are coloured by the fluctuating vertical velocity. 
   The background contour map represents the streamwise velocity fluctuation in the $x,z$-plane  at \protect\subref{fig:Q1} $y/H=0.5$ 
  and \protect\subref{fig:Q2} $y/H=0.9$. }
\end{figure}

 Figure~\ref{fig:Q1} 
 %----------
\setlength{\labelsep}{0cm}
 %----------
 shows contours of streamwise velocity fluctuations $u^\prime/U_b$ in the plane at $y/H=0.5$. Superimposed on this plot are small-scale vortical structures in the interval 
$0.5\leq y/H\leq1$ that are visualised using the normalised $Q$-criterion,
\begin{equation}
q(x,y,z,t) = {Q(x,y,z,t)} / {Q_{rms}(y)}, 
\label{eq:Qnorm}
\end{equation}
where $Q$ is the second invariant of the velocity gradient tensor \citep{Hunt1988} and $Q_{rms}$ is the root mean square of $Q$ determined using temporal and spatial averaging in $x, z$.   
In figure~\ref{fig:Q1}, it can be seen that the vast majority of the small-scale vortical structures is present inside large low-speed streaks that extend toward the surface (cf. also figure~\ref{fig:avgx_flow}).
The strong concentration of these small structures in low-speed streaks can be explained as follows. Low-speed streaks form when relatively slow moving, highly turbulent flow from the lower part of the boundary layer is ejected to the upper, less turbulent part of the boundary layer  \citep[e.g.][]{Komori1982}. This larger turbulence intensity is responsible for the concentration of small scale structures observed inside the low-speed streaks.

When the small vortices approach the surface they either align with or become  orthogonal to the surface, which are referred to by surface-aligned and surface-attached vortices, respectively. 
Figure~\ref{fig:Q2} shows an instantaneous snapshot of the upper part ($y/H\ge 0.9$) of the channel. 
Most of the surface-attached vortices were found above downwelling motions corresponding to  high-speed streaks.
As mentioned above, in low-speed streaks most of the vortical structures are present. These structures tend to align with the surface due to the shear generated underneath the divergent flow at the surface.
In these low-speed regions, the surface aligned vortices are often ring-shaped, which according to \cite{Nagaosa2003} started their life as hairpin vortices from near the wall of the channel. 

The implications of the above on interfacial mass transfer will be discussed in section  \ref{sec:flowmass}. 

%---------------------------------------------------------------------
\section{Mass transfer}
%---------------------------------------------------------------------
%---------------------------------------------------------------------
\subsection{Instantaneous results}
%---------------------------------------------------------------------
\label{sec:mass}
%----
\begin{figure}
\setlength{\labelsep}{0cm}
\centering
\begin{tabular}[t]{l l}
\sidesubfloat[]{
  \includegraphics[scale=0.122,trim={1.8cm 0 2.2cm 0},clip]{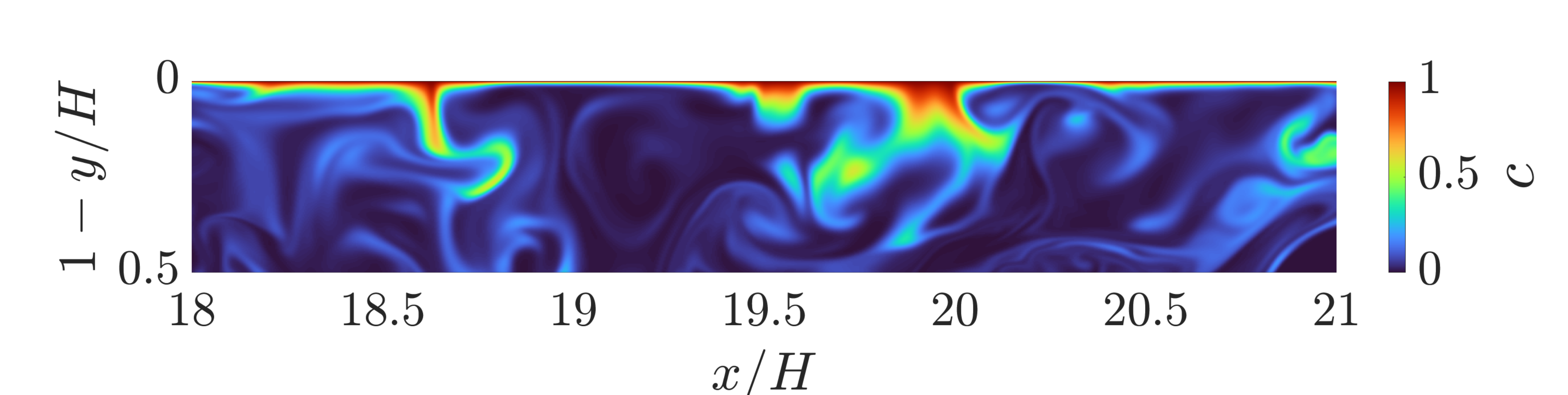}\label{fig:G08_BL_Sc7}}&
\sidesubfloat[]{
  \includegraphics[scale=0.122,trim={1.8cm 0 2.2cm 0},clip]{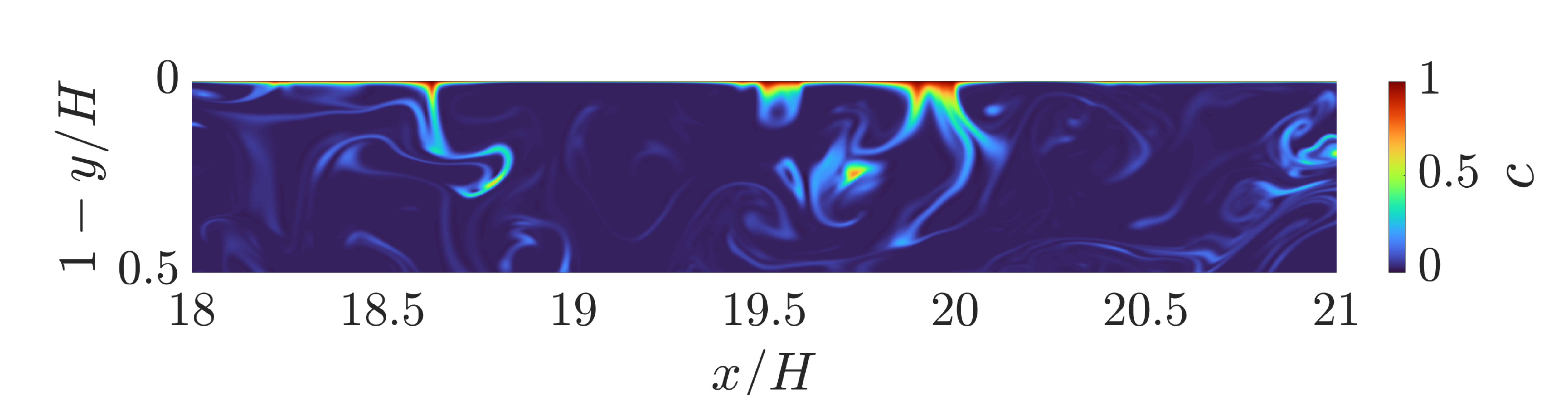}\label{fig:G08_BL_Sc100}}\\ 
\sidesubfloat[]{
  \includegraphics[scale=0.28,trim={0cm 0 .95cm 0},clip]{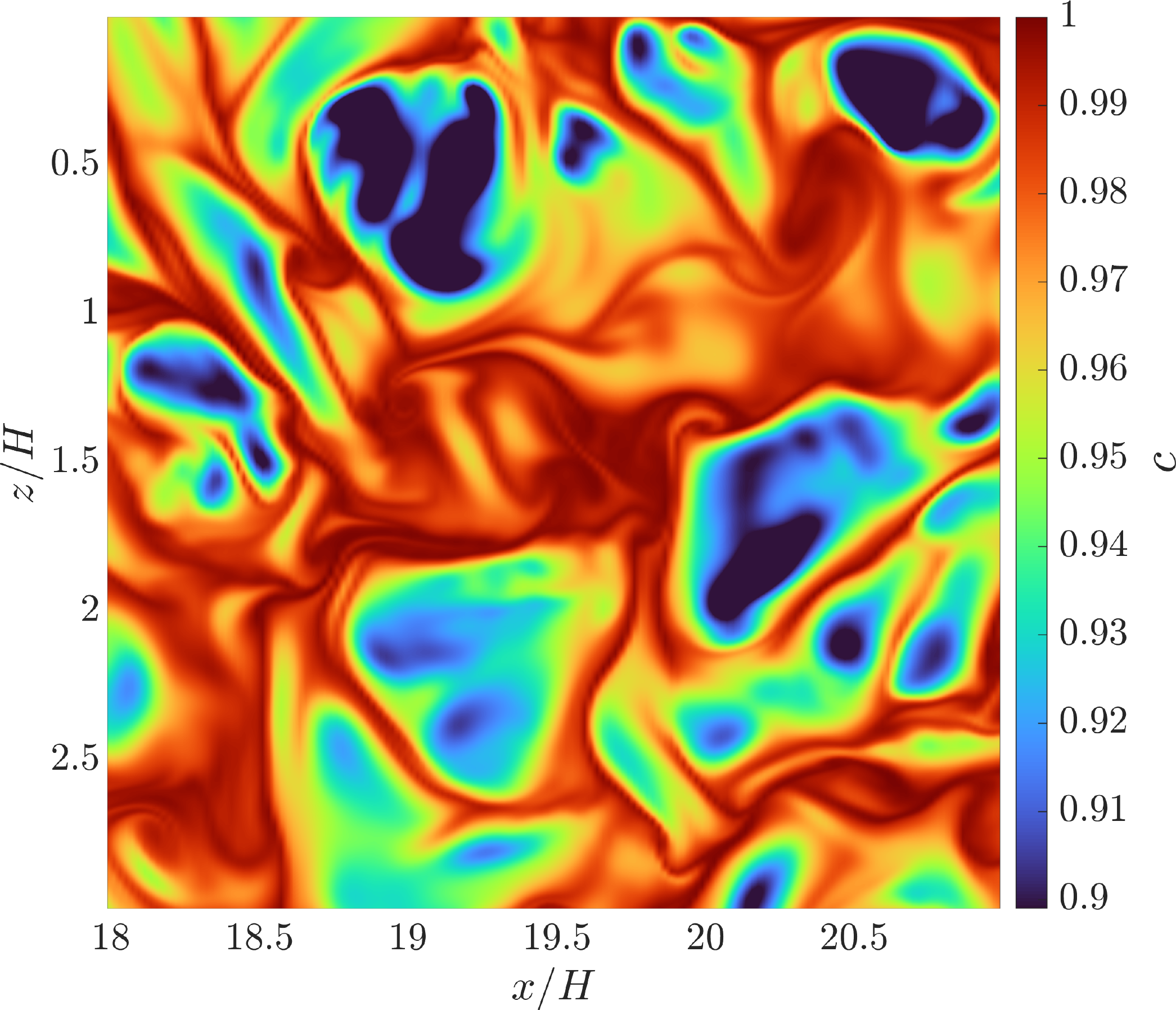}\label{fig:G08_BL_TOP7}}&
\sidesubfloat[]{
  \includegraphics[scale=0.28,trim={0cm 0 .95cm 0},clip]{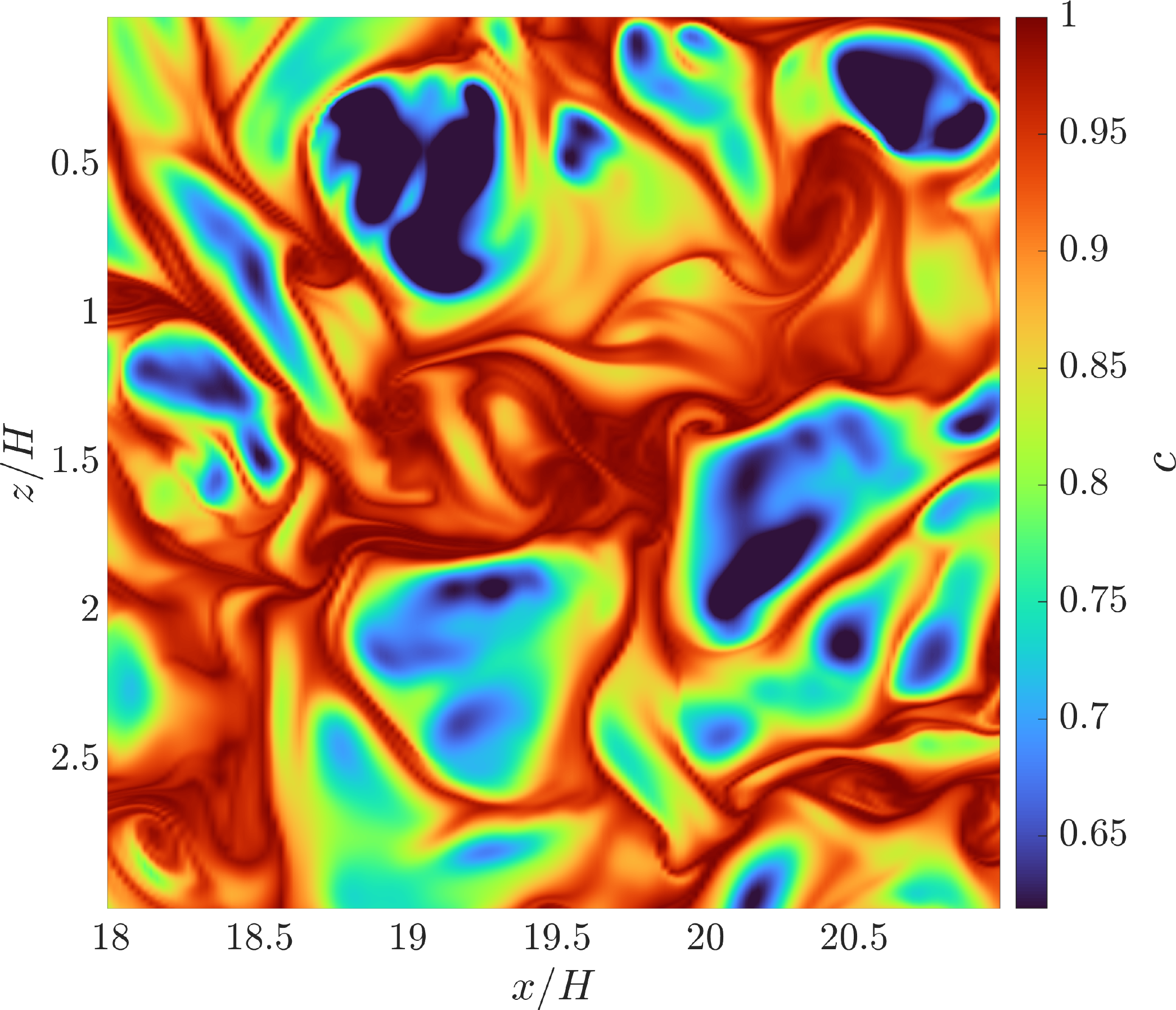}\label{fig:G08_BL_TOP100}} 
  \end{tabular}
  \caption{Typical contours of $c^*$ in the $x,y$-plane (top panes) and  
  in the $x,z$-plane at $y/H=0.9997$  (bottom panes). Shown are snapshots from G07 (at $t/t_b=42$) for  
  \protect\subref{fig:G08_BL_Sc7}, \protect\subref{fig:G08_BL_TOP7}  $Sc=7$ and 
  \protect\subref{fig:G08_BL_Sc100},\protect\subref{fig:G08_BL_TOP100} $Sc=100$. 
  Please note the different ranges in the colormaps of panes \protect\subref{fig:G08_BL_TOP7} and \protect\subref{fig:G08_BL_TOP100}.}
  \label{fig:BLqual}
\end{figure}
%-----
Figure~\ref{fig:BLqual} shows typical contours of the concentration and qualitatively depicts the interaction between the turbulent open channel flow and the scalar transport at $Sc=7$ \protect\subref{fig:G08_BL_Sc7},\protect\subref{fig:G08_BL_TOP7} and $Sc=100$ \protect\subref{fig:G08_BL_Sc100},\protect\subref{fig:G08_BL_TOP100}. 
The thickness of the concentration boundary layer $\delta$, which will be defined in equation~(\ref{eq:delta}) below,  depends on both the Reynolds and the Schmidt number.
Comparing the vertical cross-sections ($x,y$ planes) shown in figures~\ref{fig:G08_BL_Sc7} and \ref{fig:G08_BL_Sc100}, it can be clearly seen that an increase in Schmidt number at constant $Re_b$ results in much finer concentration filaments in the bulk and a significantly reduced boundary layer thickness. 
Note that for a shear-free surface, the thickness $\delta$ scales with $Sc^{-0.5}$ (cf. figure~\ref{fig:sbl1}) so that $\delta_{Sc=7} \approx \,3.78\delta_{Sc=100}$. This scaling is not taken into account in figures \ref{fig:G08_BL_TOP7} and \ref{fig:G08_BL_TOP100}, which shows surface-parallel planes at a distance of $0.0003H$ to the surface, corresponding to $\simeq 0.019\delta$ and $\simeq 0.071 \delta$ for $Sc=7$ and $100$, respectively. Hence, the range of the scalar concentrations in figure \ref{fig:G08_BL_TOP100} had to be increased by a factor of $\approx \sqrt{100/7}=3.78$ to obtain similar contours. Nevertheless, small differences can still be observed locally due to differences in diffusion in the horizontal directions.  

Note that the above reduction in $\delta$ at a fixed $Re_b$ is due to the increase in interfacial mass transfer resistance with increasing Schmidt number. At a fixed $Sc$, the increase in turbulence in the bulk associated with an increase in $Re_b$ results in improved mixing with a reduction of $\delta$, which in this case promotes mass transfer.

%---------------------------------------------------------------------
\subsection{Statistics of scalar transport}
\label{sec:scalar}
%---------------------------------------------------------------------
As mentioned in section~\ref{sec:flow} all simulations were started from fully developed turbulent flow fields with the scalars initialised by (\ref{eq:initialscalar}). After a transient period needed to ensure that the scalar statistics were quasi-steady, scalar averaging was carried out using a time-window of $\Delta t_s/t_b$ (see table 2).

\begin{figure}
  \setlength{\labelsep}{-0.6cm}
  \sidesubfloat[]{
  \includegraphics{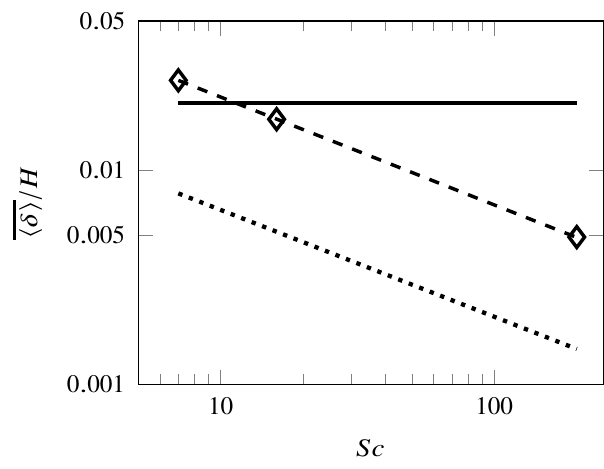}
  \label{fig:sbl1}} \quad
  \sidesubfloat[]{
  \includegraphics{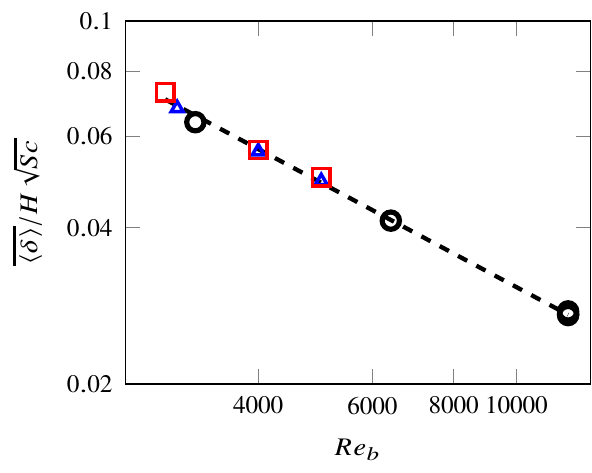}
  \label{fig:sbl2} }
  \caption{
  \protect\subref{fig:sbl1} Typical scaling of the mean boundary layer thickness $\overline{\langle\delta\rangle}/H$ with $Sc$, shown here for case G07. Also included are: \protect\myline{black,solid},~the  Kolmogorov scale ($\overline{\langle\eta\rangle}/H$);  \protect\myline{black,dotted},~the Batchelor ($\overline{\langle L_B\rangle}/H$) scale;  \protect\myline{black,dashed},~the $Sc^{-0.5}$ slope.  \protect\subref{fig:sbl2} Normalised boundary layer thickness $\overline{\langle\delta\rangle}\sqrt{Sc}/H$ as a function of $Re_b$. The symbols correspond to : \protect\mytriangle{blue},~G01-G03; \protect\mysquare{red},~G04-G06; \protect\mycircle{black},~G07-G09 and the line  \protect\myline{black,dashed} represents  $14.4Re_b^{-0.67}$. 
  \label{fig:sbl}}
\end{figure}

The thickness of the diffusive concentration boundary layer $\delta$ is identified using 
\begin{equation}
    \delta = \frac {(c_s-\overline{\langle c_b \rangle})}{ \partial c / \partial y|_{y=H}}. 
    \label{eq:delta}
\end{equation}
As illustrated in figure~\ref{fig:sbl1} for simulation G07, in all simulations $\delta$ was found to scale with $Sc^{-0.5}$ which is in agreement with the theoretical prediction for a shear-free interface \citep[e.g. ][]{ledwell:84}.
%------
%--------
Included in this plot are the thicknesses of  the Kolmogorov sublayer $\eta$ and the Batchelor sublayer $L_B$ at the interface.
Except for $Sc=7$, it was found that $L_B < \delta < \eta$, which is in agreement with \citet[][hereafter HW14]{herlina:14}. 

Figure~\ref{fig:sbl2} shows the variation of $\overline{\langle \delta \rangle} \sqrt{Sc}/H$ with the bulk Reynolds number $Re_b$. It can be seen that for $Re_b=4000$ and $5000$, the normalised boundary layer thickness is nearly independent of the computational domain size.   
The best fit through the data points was found to be  $\overline{\langle \delta \rangle} \sqrt{Sc}/H \propto Re_b^{-0.67}$. 
This will be discussed further in section \ref{sec:kl}, where the scaling will be linked to the transfer velocity. 

\begin{figure}
   \centering
   \sidesubfloat[]{\includegraphics{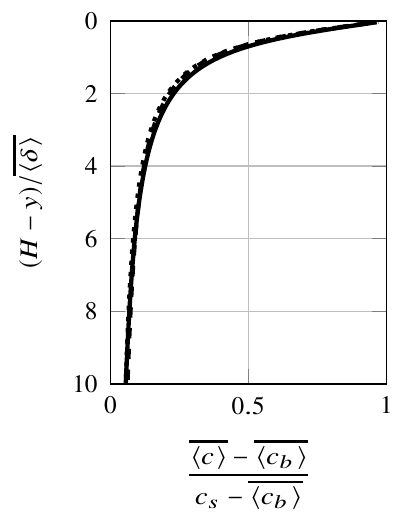} \label{fig:concRe}} \quad
   \sidesubfloat[]{\includegraphics{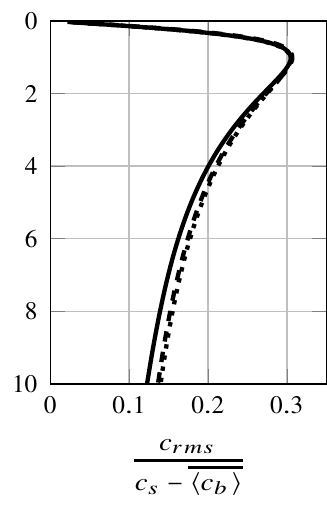} \label{fig:crmsRe}}\quad
   \sidesubfloat[]{\includegraphics{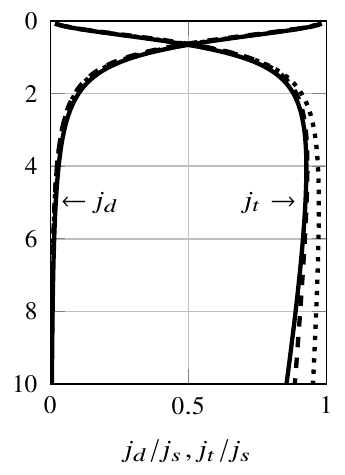} \label{fig:flux_scalRe1}} 
   \caption{
     Vertical profiles of normalised \protect\subref{fig:concRe} mean concentration, \protect\subref{fig:crmsRe} r.m.s. of concentration, \protect\subref{fig:flux_scalRe1} mean diffusive $j_d/j_s$ and turbulent $j_t/j_s$ mass fluxes, where $j_s$ denotes the surface mass flux. 
    Shown are profiles at $Sc=16$ for various bulk Reynolds numbers~: $Re_b=3200$ (G07, solid lines),  $Re_b=6400$ (G08, dashed lines) and $Re_b=12000$ (G09, dotted lines).  
    }
    \label{fig:scalstatRe}
\end{figure}

Figure~\ref{fig:scalstatRe} shows normalised mean vertical profiles of the concentrations, the r.m.s. of concentration fluctuations and the mass fluxes at various Reynolds numbers (G07, G08 and G09). As discussed above, the boundary layer thickness $\delta$ depends on both the molecular diffusivity ($Sc$) and the Reynolds number ($Re_b$). 
Thus, it is expected that the vertical profiles of the normalised mean scalar quantities exhibit self-similarity when the vertical $(H-y)$ direction is normalised by $\overline{\langle \delta \rangle}$. (Note that the profiles for the different Schmidt numbers also collapse). 
%------
%-------

In all simulations, the magnitude of $\overline{\langle \delta \rangle} /H$ was found to be virtually identical to the distance between the surface and the location at which the r.m.s. of the concentrations 
\begin{equation}
c_{rms} = \sqrt{\overline{\langle c^2 \rangle - \langle c \rangle^2}}
\end{equation}
reaches its maximum. 
Hence, the peak in figure~\ref{fig:crmsRe} is located at $(H-y)/\overline{\langle \delta \rangle}=1$. The maximum $c_{rms}/(c_s-c_b)$ values were $\approx 0.3$, which is in agreement with previous numerical \citep{Magnaudet2006,herlina:14,herlina:19} and experimental \citep{Atmane2002} results.  
The lower normalised $c_{rms}$ peak values of  $\approx0.1-0.2$ obtained in the experiments of  \cite{herlina:08,Janzen2010} indicate a partially contaminated surface, as confirmed by the numerical studies of \cite{Khakpour2011,Wissink2017}.

The total averaged vertical mass flux, $j=j_d +j_t$, comprises a diffusive component
 \begin{equation}
 j_d(y)=-D  \frac{\partial \overline{\langle c \rangle}}{\partial {y}}
\end{equation}
and a turbulent component 
 \begin{equation}
 j_t(y)=\overline{\langle c^\prime v^\prime \rangle}.
\end{equation}
Figure~\ref{fig:flux_scalRe1} illustrates that $j_d$ dominates at the surface, as $v'$ is damped due to the two-dimensionality imposed by the free-slip boundary condition. With increasing distance to the surface the contribution of $j_d$ to the total mass flux reduces, while at the same time the contribution of $j_t$ becomes increasingly important until it entirely dominates $j$. It can also be seen that at $(H-y)/\overline{\langle \delta \rangle}\simeq0.65$ the diffusive and turbulent mass fluxes become equal. Furthermore, it was found that in the range $2\lessapprox(H-y)/\overline{\langle \delta \rangle}\lessapprox10$, the turbulent mass flux $\overline{\langle c'v' \rangle}$ agrees reasonably well with the mass flux at the surface ($j_s$).

From the surface ($y=H$) down to a depth of $(H-y)\approx 2\overline{\langle \delta \rangle}$, the normalised mean vertical profiles of the scalar quantities  shown in figure~\ref{fig:scalstatRe} were found to nearly collapse with the profiles reported in \citet[][hereafter HW19]{herlina:19} (not included in the plot), where the interfacial mass transfer was driven by isotropic-turbulence diffusing upwards from the opposite boundary. For the highest Reynolds number case (G09), the agreement between the open channel profiles produced here and in the case driven by isotropic turbulence remains reasonably well up to a distance of at least $10\overline{\langle \delta \rangle}$ from the free-surface. This indicates that the strong anisotropy present in the velocity field of turbulent open channel flow (which increases with increasing Reynolds number) does not affect the shape of the normalised profiles at least up to the aforementioned depth.

%---------------------------------------------------------------------
\subsection{Scaling of mass transfer velocity}
%---------------------------------------------------------------------
\label{sec:kl}
The local, instantaneous transfer velocity $k_l$ is defined as 
\begin{equation} 
k_l (x,z,t) =\left| {\frac{-D\left. {\partial c(x,z,t)}/{\partial y} \right |_{y=H} }
                          { c_s- \langle c_b \rangle }}
                     \right|.
\end{equation} 
For open channel flow with no wind-shear, the mean transfer velocity 
\begin{equation}
K_L   =  \left| \frac{\left. j_d\right |_{y=H}}{c_s-\overline{\langle c_b \rangle}}\right|
= \langle \overline{k_l(x,z,t)} \rangle 
\end{equation} 
is commonly parameterised using the bulk Reynolds number $Re_b$ or the friction Reynolds number $Re_\tau$.  
In figure~\ref{fig:sbl} it was found that $\overline{\langle \delta \rangle/H}$ scales with $Sc^{-0.5}$ and $Re_b^{-0.67}$. 
Hence, as 
\begin{equation}
\overline{\langle \delta \rangle} =D/K_L, 
\end{equation}
the normalised transfer velocities $K_L/U_b$ obtained here scale with $Sc^{-0.5}$ and $Re_b^{-0.33}$.  
The former scaling was found to be valid only for clean (surfactant-free) surfaces \citep{Wissink2017}. The power $n=-0.33$ in the latter scaling is somewhat smaller than the one found in the DNS of \citet[][hereafter NH12]{Nagaosa2012}, who obtained $K_L \propto Re_b^{-0.25}$ in small-box simulations at $Sc=1$. 
%------------
\begin{figure}
  \centering
  \sidesubfloat[]{\includegraphics{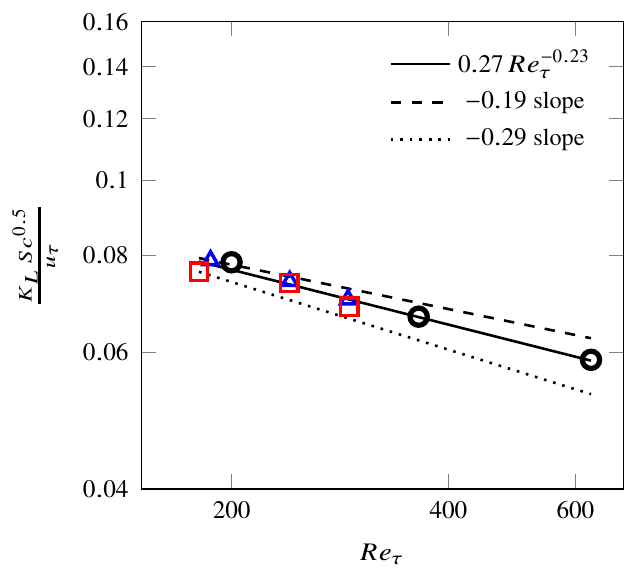} \label{fig:klRe}}\quad
  \sidesubfloat[]{\includegraphics{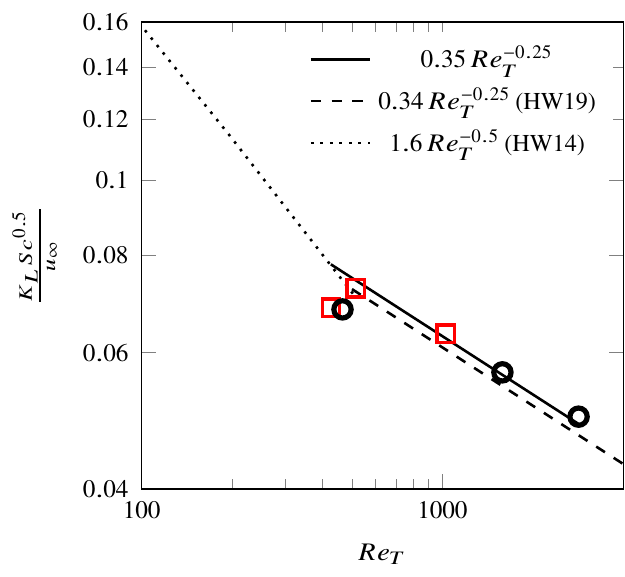} \label{fig:klReT}}
   \caption{Normalised mass transfer velocity as a function of \protect\subref{fig:klRe} $Re_\tau$ and \protect\subref{fig:klReT} $Re_T$, compared with the fitted-relations found in HW14 and HW19. The symbols denote : \protect\mytriangle{blue},~G01-G03; \protect\mysquare{red},~G04-G06; \protect\mycircle{black},~G07-G09.   }
  \label{fig:Re}
\end{figure}
%------------
%
When using the friction Reynolds number $Re_\tau$ rather than $Re_b$, $K_L/u_\tau$ was found to scale with $Re_\tau^{-0.23}$, see figure~\ref{fig:klRe}. This power of $n=-0.23$ is similar to the powers  found in open channel flow experiments of  \cite{Moog1999, Lau1975, Gulliver1989} where $n$ ranges between $-0.19$ and $-0.29$.  

\cite{Theofanous1976} proposed to use the turbulent Reynolds number $Re_T$, defined by 
\begin{equation}
Re_T = u_\infty 2L_\infty/\nu, 
\label{eq:ReT}
\end{equation}
as a measure of turbulence characteristics that is independent on the way that turbulence is generated, e.g. by wind-shear, bottom-shear, or buoyancy. In (\ref{eq:ReT}), $u_\infty$ is the turbulence intensity and $L_\infty$ is the streamwise integral length scale, taken here as $u_{rms}$ and $L^x_{uu}$ evaluated at the edge of the surface influenced layer $y=y_\infty$, see table~\ref{tab:ReT}. Note that $Re_T$ could not be computed for the smallest domain size since no decorrelation of $u$ in the streamwise direction was achieved. As can be seen in table~\ref{tab:ReT}, our data fall near and above the critical turbulent Reynolds number $Re_T=500$, that in the two-regime model of \cite{Theofanous1976} separates the large-eddy ($Re_T\leq500$) from the small-eddy ($Re_T\geq500$) regime. 
Figure~\ref{fig:klReT} shows that in our open channel flow simulations $K_L Sc^{0.5}/u_\infty$ scales with $0.35Re_T^{-0.25}$ when $Re_T> 500$. The normalised $K_L/u_\infty$ values and hence its scaling with $Re_T$ are in close agreement with the results found in the absence of mean shear by HW19, which confirms that for $Re_T>500$ the turbulent characteristics are dominated by small eddies (i.e. $K_L Sc^{-0.5}/u_\infty \propto Re_T^{-0.25}$), also in open channel flow. 
\begin{table}
  \begin{center}
  \rule{\textwidth}{.5pt}\vspace{2ex}
  \def~{\hphantom{0}}
  \small
  \begin{tabular}{c c c c c c c c c c c c c c c c }
    & G01 & G02 & G03 & G04 & G05 & G06 & G07 & G08 & G09\\[3pt]
    $y_\infty/H$ & 0.7034  &  0.7397  &  0.7573  &  0.6666  &  0.7102  &  0.7208  &  0.6327  &  0.6651  &  0.7226\\
    $u_\infty/U_b$ & 0.0711 & 0.0663 & 0.0589 & 0.0696  &  0.0610  &  0.0631  &  0.0718  &  0.0669  &  0.0621\\
    $L_\infty/H$ & - & - & - & 1.0626  &  1.0550  &  1.6246  &  1.0121  &  1.8457  &  1.8992 \\
    $Re_T$ & - & - & - & 425 & 515 & 1025 & 465 & 1581 & 2833 \\
    $\Delta t_s/t_b$ & 200 & 100 & 120 & 230 & 130 & 120 & 60 & 60 & 60\\
  \end{tabular}
  \end{center}
   \caption{Parameters used in the definition of the turbulent Reynolds number for the simulations listed in Table~\ref{tab:mass}. 
  The location $y_\infty$ identifies the edge of the surface influenced layer (cf. section \ref{sec:flow}),  
  $L_{\infty}, u_{\infty}, Re_T$ are defined in equation (\ref{eq:ReT}).
  Note that all values were obtained by averaging over a time-window $\Delta t_s/t_b$ in which the scalar statistics are quasi-steady, see table~\ref{tab:mass}.\\[-.5ex]
  \noindent\hspace{-30pt}\rule{\textwidth}{.5pt}  }
  \label{tab:ReT}
 \end{table}
%------------------------------------------------------
\subsection{Flow structures and mass transfer}
\label{sec:flowmass}
%------------------------------------------------------
\begin{figure}
  \includegraphics{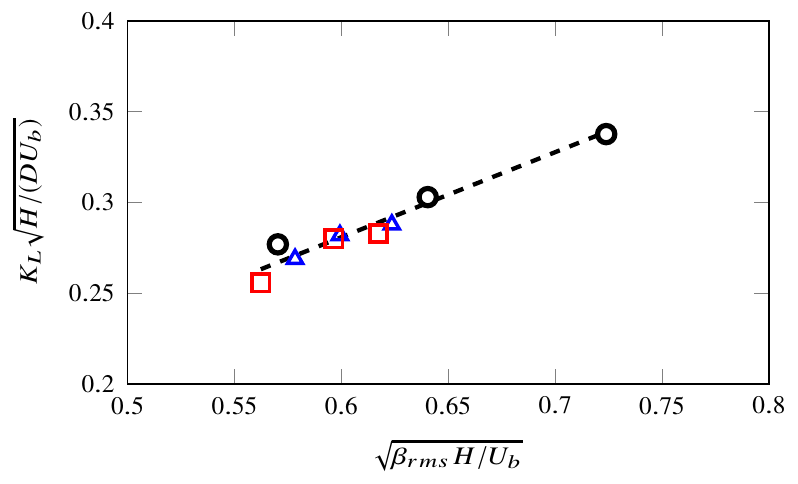}
  \caption{Variation of $K_L \sqrt{H/(D U_b)}$ with $\sqrt{\beta_{rms}H /U_b}$. The coefficient of proportionality of the fitted-dashed line was found to be $0.47$. The symbols indicate : \protect\mytriangle{blue},~G01-G03; \protect\mysquare{red},~G04-G06; \protect\mycircle{black},~G07-G09.}
  \label{fig:SDM}
\end{figure}
%----------
\begin{figure}
  \setlength{\labelsep}{-.1cm}
  \sidesubfloat[]{\includegraphics{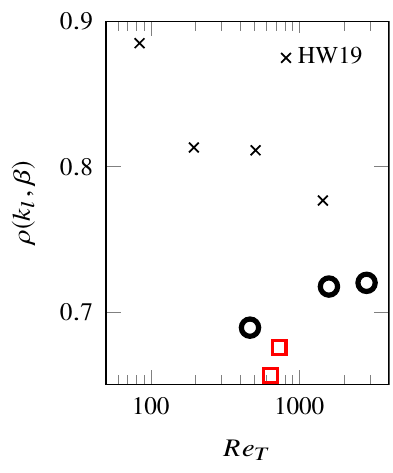} \label{fig:rhoklbetaRT}}\,\,
  \sidesubfloat[]{\includegraphics{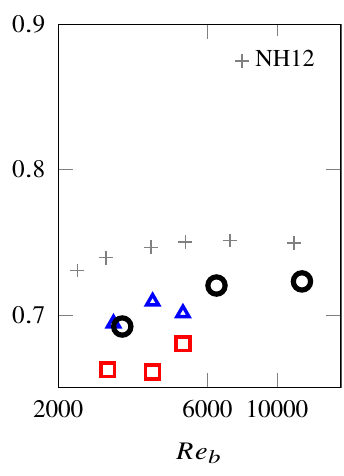} \label{fig:rhoklbetaRb} }\,\,
  \sidesubfloat[]{\includegraphics{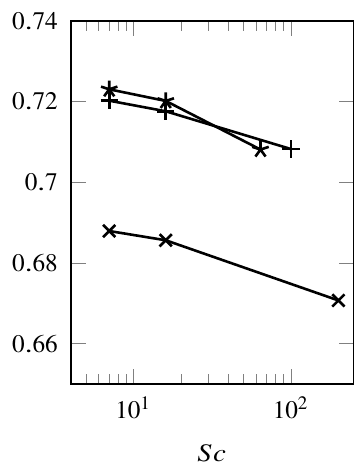} \label{fig:rhoSc}}
   \caption{Correlation between the instantaneous mass transfer velocity $k_l$ and surface divergence $\beta$  as a function of 
   \protect\subref{fig:rhoklbetaRT} $Re_T$, 
   \protect\subref{fig:rhoklbetaRb} $Re_b$ and 
   \protect\subref{fig:rhoSc} $Sc$. In \protect\subref{fig:rhoklbetaRT} data from G05-G09 at $Sc=16$ are compared with HW19 at $Sc=20$. In \protect\subref{fig:rhoklbetaRb} data from G01-G03 at $Sc=8$ and G04-G09 at $Sc=7$ are compared with  NH12 at $Sc=1$. In \protect\subref{fig:rhoSc} data are from G07-G09. The symbols correspond to~: 
   \protect\mytriangle{blue},~G01-G03; 
   \protect\mysquare{red},~G04-G06;  
   \protect\mycircle{black},~G07-G09; while the lines represent~:  
   \protect\mylinsym{$\times$},~G07; 
   \protect\mylinsym{$+$},~G08; 
   \protect\mylinsym{$\ast$},~G09. \label{fig:rhoklbeta1}}
\end{figure}
%----------
%
Previous authors have shown that the surface divergence model of \cite{McCready1986}, 
$K_L\propto \sqrt{\beta_{rms}D}$, where $\beta_{rms}$ is the r.m.s of the surface divergence 
$\left( \beta= \left. -{\partial v^\prime}/{\partial z} \right |_{z=H}
\right)$,  
 generally provides a good prediction of  the mass transfer velocity, in open channel flow \citep{Sanjou2017, Turney2013} as well as in turbulence without mean shear \citep[HW19, ][]{McKenna2004}. 
The approximately linear variation of $K_L\sqrt{H/(DU_b)}$ with $\sqrt{\beta_{rms}H /U_b}$, shown in figure~\ref{fig:SDM}, confirms that the surface divergence model performs reasonably well also in the present simulations. 
For interfacial mass transfer driven by isotropic turbulence, it was found that despite the good correlation between $\beta_{rms}$ and the mean transfer velocity $K_L$, the correlation between the local, instantaneous transfer velocity $k_l (x,z,t)$ and the instantaneous surface divergence $\beta$ tends to deteriorate with $Re_T$ \citep{herlina:19}. 
As explained below, in our present simulations no such deterioration of the time-averaged correlation 
$\overline{\rho(k_l(x,z,t),\beta(x,z,t))}$ with an increase in turbulence level was observed. Hereafter, the aforementioned correlation will be referred to as $\rho(k_l, \beta)$.

Figure~\ref{fig:rhoklbetaRT} depicts the variation of $\rho(k_l, \beta)$ with $Re_T$ obtained in the present simulations, combined with results from HW19. 
It can be seen that initially, for isotropic turbulence ($Sc=20$) the correlation $\rho(k_l, \beta)$ tends to reduce with increasing $Re_T$ and appears to reach a plateau around $\rho \approx 0.8$. 
In contrast, in the large-box simulations of the present open channel flow study at $Sc=16$,  the correlation $\rho(k_l, \beta)$ approaches a somewhat smaller value of approximately $0.72$ from below. As can be seen in figure ~\ref{fig:rhoklbetaRb}, a similar improvement of $\rho(k_l,\beta)$ with $Re_b$ was also observed in the present small-box simulations and, somewhat less clear,  also in the mid-sized simulations. Note that this is in agreement with the trend observed in small-box simulations reported in NH12 at $Sc=1$. The good agreement achieved between the small and large box simulations is likely due to the incomplete de-correlation in the streamwise direction in the small box, which mimics the presence of  large streamwise structures as can only be fully captured in much larger computational domain. Contrastingly, in the mid-sized box a marginal streamwise de-correlation was achieved causing the maximum size of the streamwise structures to become significantly smaller than in the large box simulations. Note that a more detailed study of the effect of the domain size on mass transfer is beyond the scope of this paper.

The Schmidt number effect on the correlation coefficient $\rho(k_l,\beta)$ is shown in figure~\ref{fig:rhoSc}. The correlation was found to decrease with increasing Schmidt number. 
This can also be seen in figure~\ref{fig:rhoklbetaRb} when comparing the present results with NH12. For $Sc = 1$ momentum and scalar diffusivity are identical and the correlation is not influenced by differences in diffusive time scales. With increasing scalar diffusivity, the difference in diffusive time scales between momentum and scalar increases, resulting in a reduced correlation between $\beta$ and $k_l$.  
\begin{table}
  \begin{center}
  \rule{\textwidth}{.5pt}\vspace{2ex}
  \def~{\hphantom{0}}
  \small
  \begin{tabular}{c c c c c c c c c c c c c c c c }
    Run 
    & \multicolumn{2}{c}{low-speed}
    & \multicolumn{2}{c}{high-speed} &
     \\
     & $\rho(k_l,\beta)$  & $A$
    & $\rho(k_l,\beta)$  & $A$
    & $\rho(k_l,\beta)_{tot}$\\[3pt]
    G07 & 0.761 & 0.292 & 0.592 & 0.314 & 0.688 \\
    G08 & 0.773 & 0.291 & 0.647 & 0.325 & 0.720\\
    G09 & 0.773 & 0.291 & 0.660  & 0.319 & 0.722
  \end{tabular}
  \caption{Correlation coefficient, $\rho(k_l,\beta)$, at $Sc=16$ for $Re_b=3200$ (G07), $Re_b=6400$ (G08) and $Re_b=12000$ (G09), together with the corresponding fractions of the free-surface area $A$ covered by high ($u > \langle {u} \rangle + 0.5\sigma(u)$) and low ($u < \langle{u} \rangle - 0.5\sigma(u)$) speed regions.\\[-.5ex]
  \noindent\hspace{-30pt}\rule{\textwidth}{.5pt}} 
  \label{tab:corr_areas}
  \end{center}
\end{table} 
%------
To investigate the slight increase of $\rho(k_l, \beta)$ with the Reynolds number - observed in open channel flow simulations - in more detail, table~\ref{tab:corr_areas} reports the correlation $\rho(k_l,\beta)$ separately for low-speed regions ($u <  \langle u \rangle - 0.5\sigma (u)$) and high-speed regions ($u >  \langle u \rangle + 0.5\sigma (u)$), where $\sigma (u)$ denotes the standard deviation of $u$. 
It was found that $\rho(k_l,\beta)$ is approximately independent of $Re_T$ in the low-speed regions, while 
it increases with $Re_T$ in the high-speed regions. As a result, an overall increase in the correlation was observed with increasing turbulence Reynolds number.  
This increased correlation was linked to the distribution of surface parallel vortical structures (SPVS) close to the surface. 
These structures are responsible for the vertical mixing of fluid close to the surface and are relatively stable. In an idealised case, they would look like a surface parallel ring vortex that continuously brings up unsaturated fluid in its centre to the surface leading to local maxima in $k_l$ and $\beta$, while on its outer side the now partially saturated fluid is returned to the bulk leading to local minima in $k_l$ and $\beta$ as illustrated in figures~\ref{fig:snpst3}a and c. Hence, these SPVS typically contribute to a good correlation $\rho(k_l,\beta)$. 
For low $Re_T$, these SPVS are mainly present in the low-speed regions, while for larger $Re_T$ this distribution was observed to become more uniform (possibly due to increased mixing), see figures~\ref{fig:snpst3}b and d. As a result, $\rho(k_l,\beta)$ was found to increase inside the high-speed regions, and hence overall.
The above is consistent with the observation that $\rho(k_l,\beta)$ was found to be markedly higher in the low-speed regions than in the high-speed regions.

%-------
\begin{figure}
\setlength{\labelsep}{-1cm}
  \sidesubfloat[]{
  \includegraphics[width=1\textwidth,trim={0cm 0cm 0cm 0cm},clip]
   {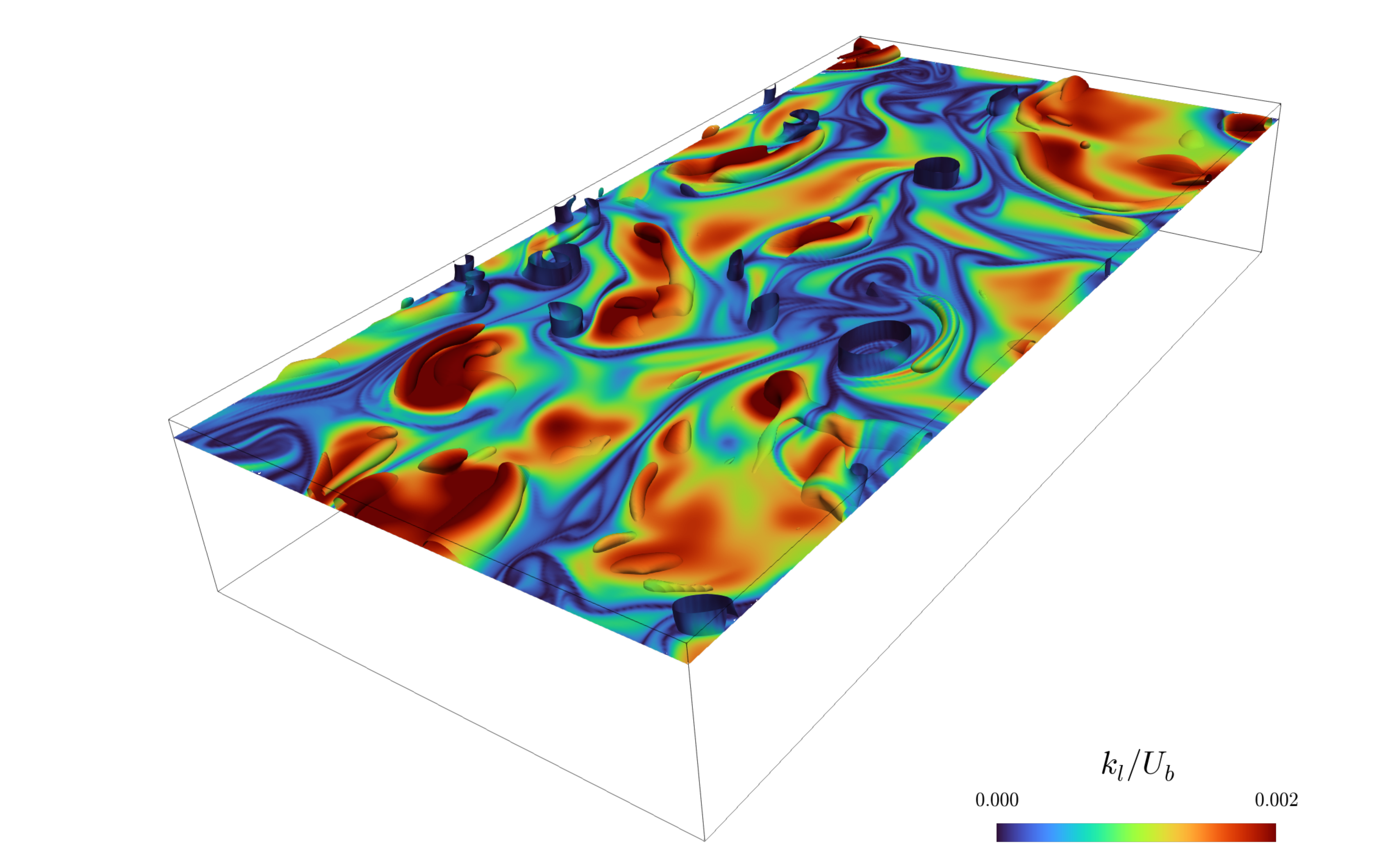}
   \label{fig:klG07}
  }\\
  \sidesubfloat[]{
  \includegraphics[width=1\textwidth,trim={0cm 0cm 0cm 0cm},clip]
  {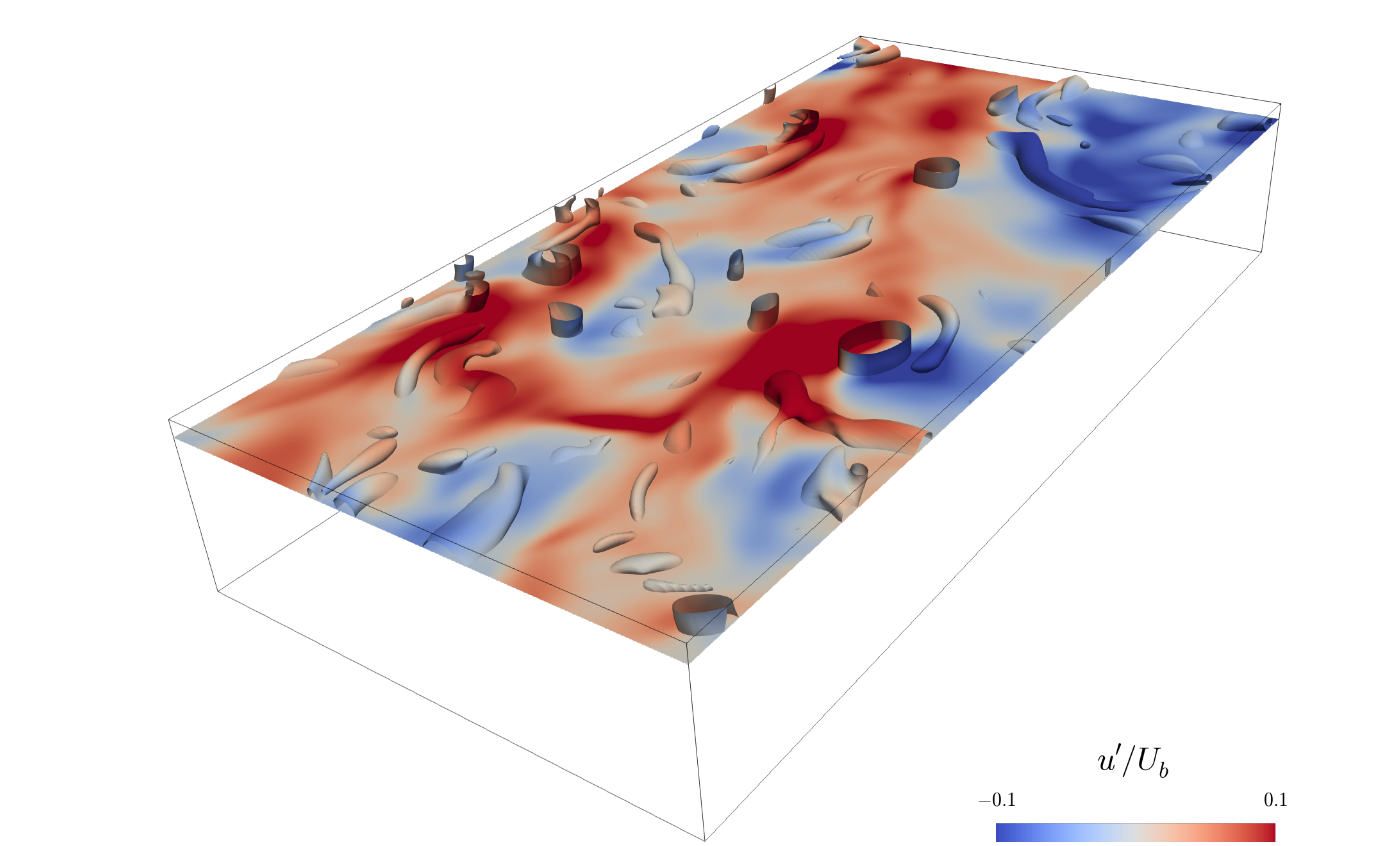}
  \label{fig:uG07}
  }
  \caption{For caption see next page.}
  \end{figure}
  \begin{figure}
  \ContinuedFloat
  \sidesubfloat[]{
  \includegraphics[width=1\textwidth,trim={0cm 0cm 0cm 0cm},clip]
  {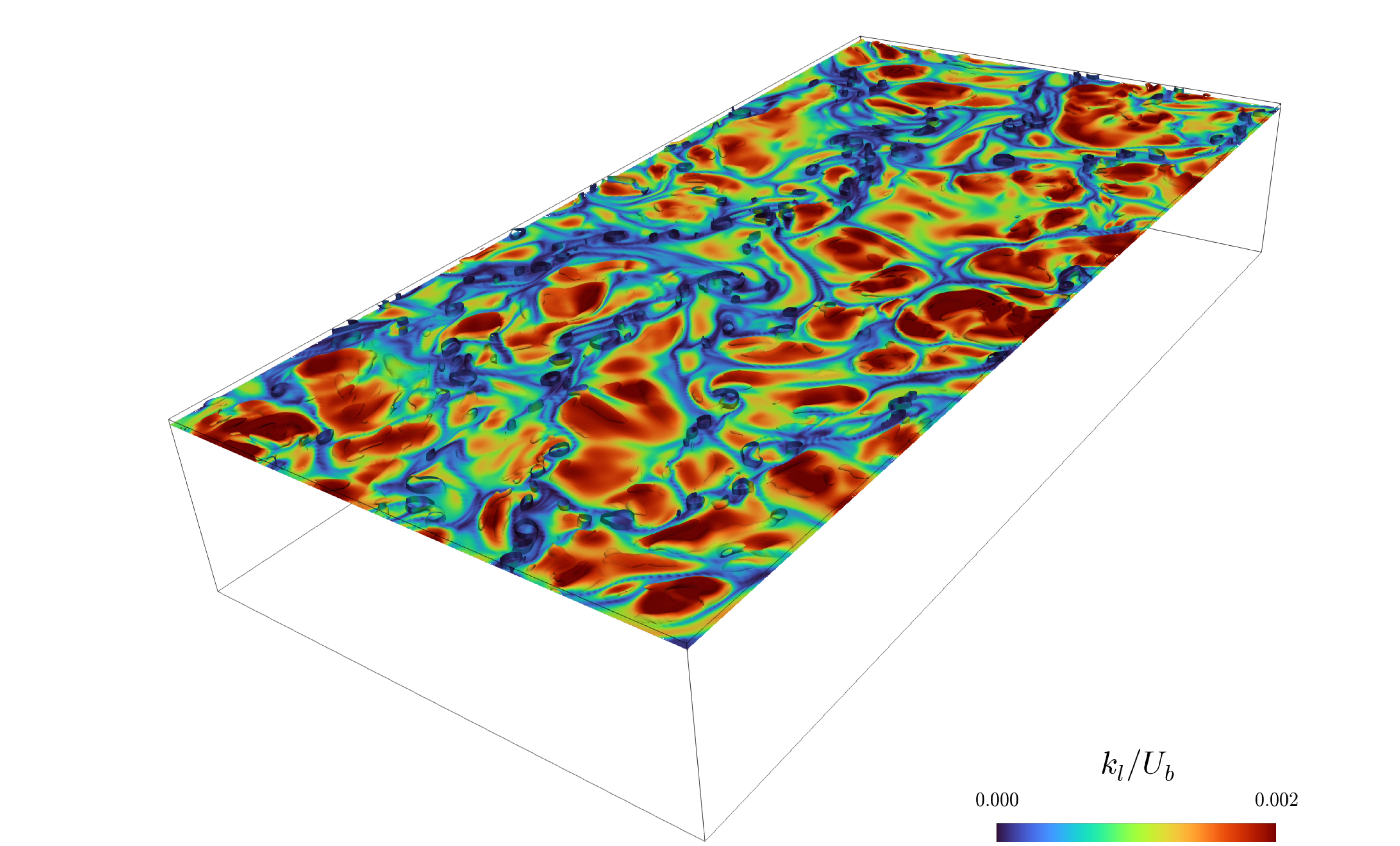}
  \label{fig:klG09}
  }\\
  \sidesubfloat[]{
  \includegraphics[width=1\textwidth,trim={0cm 0cm 0cm 0cm},clip]
  {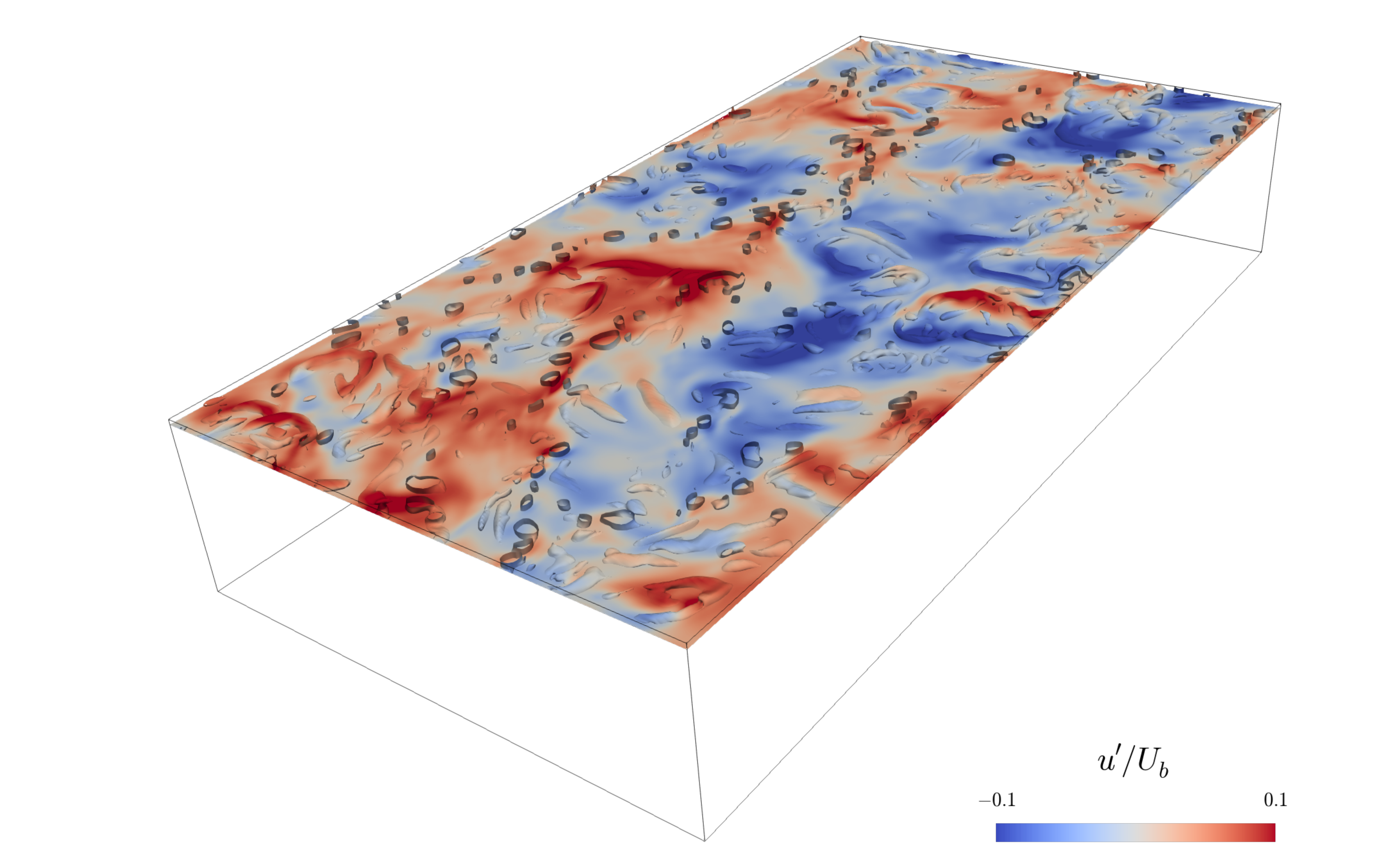}
  \label{fig:uG09}
  }\\
  \caption{(see also previous page.) Typical 3D pairs of concurrent snapshots : \protect\subref{fig:klG07}, \protect\subref{fig:uG07} case G07 at $Sc=200$, $Re_b=3200$ and \protect\subref{fig:klG09}, \protect\subref{fig:uG09} case G09 at $Sc=64$, $Re_b=12000$. The structures are visualised by iso-surfaces of the Q-criterion at $q=0.1$. Both the background $x,z$-plane (located at $y=0.9H$ for G07, $y=0.97$ for G09) and the vortical structures are coloured by $k_l$ in \protect\subref{fig:klG07},\protect\subref{fig:klG09} and by $u^\prime$ in \protect\subref{fig:uG07},\protect\subref{fig:uG09}.} 
  \label{fig:snpst3}
\end{figure}
%-------

\begin{figure}
  \includegraphics{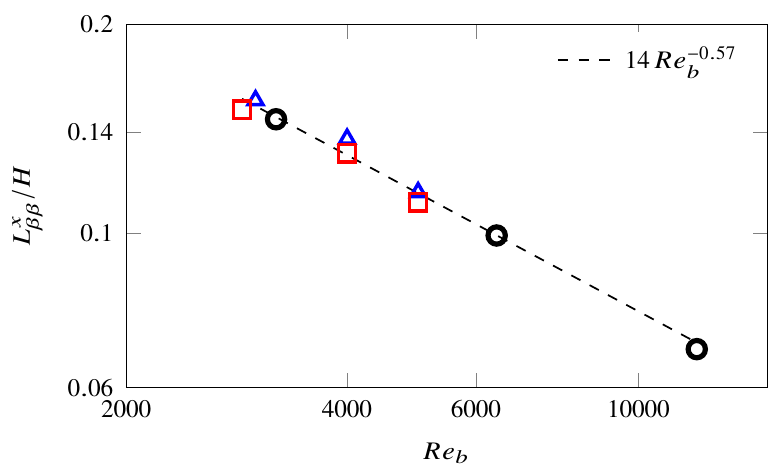}
  \caption{Streamwise integral length scale of the surface divergence as a function of the bulk Reynolds number. The symbols represent : \protect\mytriangle{blue},~G01-G03; \protect\mysquare{red},~G04-G06; \protect\mycircle{black},~G07-G09.}
  \label{fig:Lbeta}
\end{figure}

By comparing figures~\ref{fig:snpst3}a and \ref{fig:snpst3}c it can also be seen that the typical size of the $k_l$ pattern (and hence the surface divergence pattern) reduces with increasing  Reynolds number. This is consistent with the results shown in figure~\ref{fig:Lbeta}, where the integral length scale $L^x_{\beta\beta}$ was found to decrease with increasing $Re_b$.
%-----------
%----------

\begin{figure}
  \includegraphics{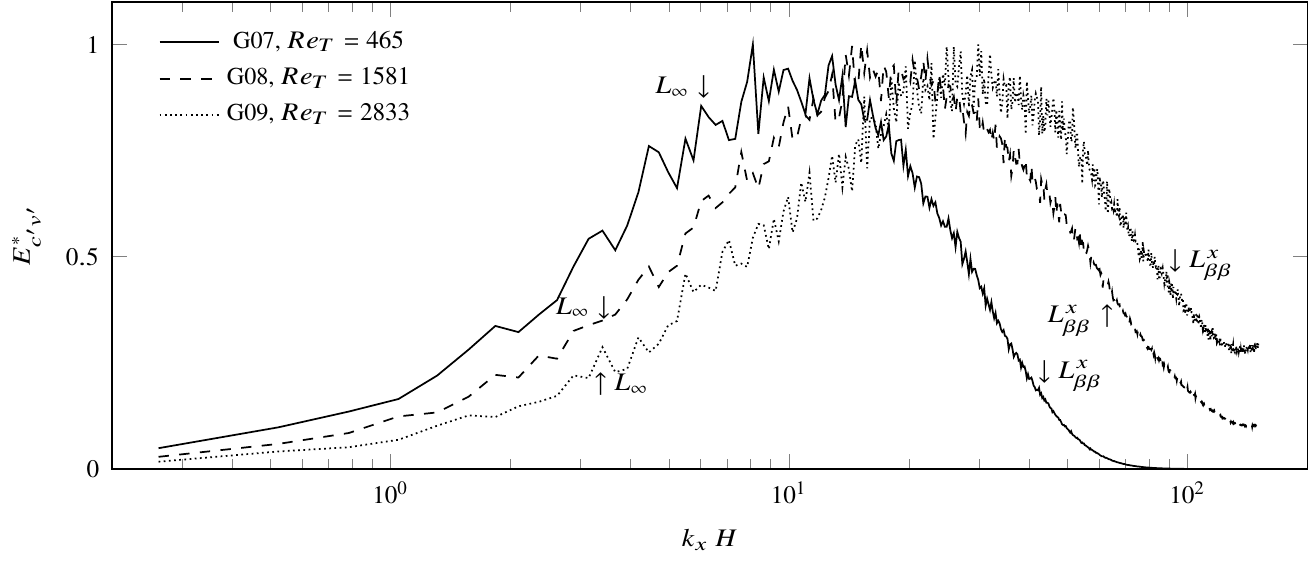}
  \caption{Normalised pre-multiplied spectral density of the turbulent mass flux $E^*_{c'v'}=k_x\,E_{c'v'}/(k_x\,E_{c'v'})_{max})$ at $(H-y)/\delta=5$, from simulations G07-G09.}
  \label{fig:spectra_tot}
\end{figure}

Because of the reasonably good correlation between surface divergence and mass transfer, it is inferred that turbulence structures of similar size as the integral length scale of the surface divergence $L^x_{\beta\beta}$ are of key importance to the overall mass transfer. Additionally, above it was found that $K_L/u_\infty \propto Re_T^{-0.25}$, which indicates that mass transfer is significantly influenced by very small scale turbulent structures. 
To explore this further, first the importance on the vertical mass transfer of turbulence structures of various size is investigated in figure~\ref{fig:spectra_tot}. Here, the normalised premultiplied spectrum  of $c^\prime v^\prime$ is shown as a function of the normalised wavenumber $k_x H$. Also indicated in the figure are the locations of the wavenumber   associated with $L_\infty$ ($k_\infty=2\pi/L_\infty$) and the wavenumber associated with $L^x_{\beta\beta}$ ($k_\beta=2\pi/L^x_{\beta\beta}$). It can be seen that in all three cases, $L^x_{\beta\beta}$ is associated with much smaller turbulence structures (higher wavenumbers) than $L_\infty$. The peaks in the premultiplied energy spectra were located in between the wavenumbers $k_\infty$ and  $k_\beta$. Also, as expected, the peaks tend to move towards larger wavenumbers (smaller scales) at higher turbulence levels (from $k_x H \approx 10$ for $Re_T=465$ to $\approx 20$ for $Re_T=1581$ to $\approx 30$ for $Re_T=2833$).  For the cases $Re_T>500$, the peaks were located at wavenumbers significantly larger than $k_\infty$.

\begin{figure}
   \setlength{\labelsep}{-0.2cm}
   \sidesubfloat[]{\includegraphics{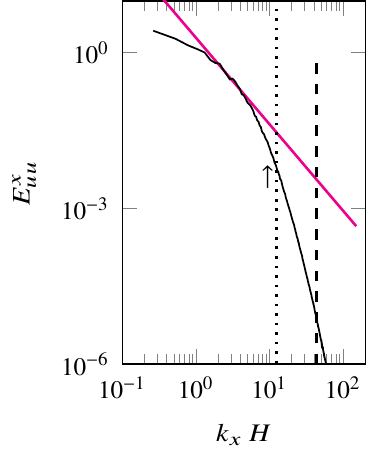} \label{fig:EuuG07}} \quad
   \sidesubfloat[]{\includegraphics{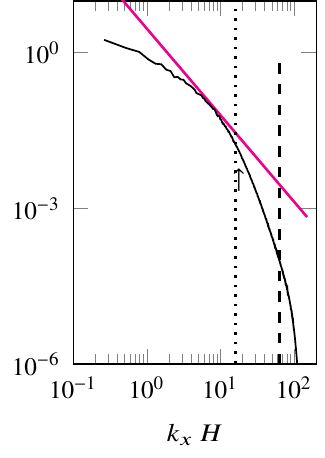} \label{fig:EuuG08}} \quad
   \sidesubfloat[]{\includegraphics{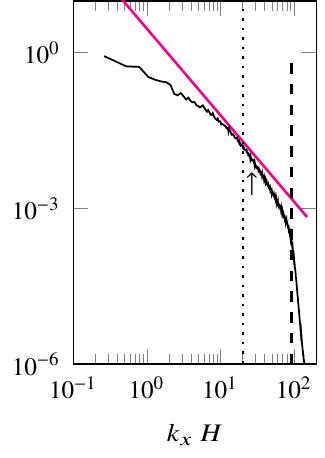} \label{fig:EuuG09}}
   \caption{Streamwise 1D-turbulent energy spectrum ($E_{uu}^x$) from simulations \protect\subref{fig:EuuG07}~G07, \protect\subref{fig:EuuG08}~G08 and \protect\subref{fig:EuuG09}~G09 at $y=y_\infty$ (cf. Table~\ref{tab:mass}). The arrows indicate the locations of $k_x$ associated with the approximate peaks in $k_xE^*_{c'v'}$, while the lines indicate : 
   \protect\myline{magenta,solid},~${-5/3}$~slope;  
   \protect\myline{black,dashed},~location of $k_\beta=2\pi/L_{\beta\beta}^x$;
   \protect\myline{black,dotted},~location of $k_{\lambda_T}=2\pi/L_{\lambda_T}^x$.}
   \label{fig:Euu_beta}
\end{figure}

Second, in figure~\ref{fig:Euu_beta}, for each simulation the approximate location of the aforementioned peaks in the streamwise 1D turbulent energy spectrum is shown. It can be seen that the wavenumber associated with each peak is located close to the Taylor microscale ($\lambda^x_T$), where dissipative effects are no longer negligible.

Third, also shown in figure~\ref{fig:Euu_beta} are the exact locations of $k_\beta$. 
In each of the simulations, $k_\beta$ was found to be markedly larger than the wavenumber associated with the Taylor microscale $k_{\lambda T}$, and hence small scale turbulence dissipation significantly affects surface divergence. This is in agreement with the theoretical considerations of \cite{McCready1986} and further evidences the association of the surface divergence integral length scale ($L^x_{\beta\beta}$) with small rather than large turbulence scales. 
The above confirms the importance of small (dissipative) scales on interfacial mass transfer in turbulent open channel flow for  $Re_T$ larger than $\approx 500$.
%----------
%----------
%----------

%-------------------------------------------------------------------------
\section{Conclusions}
%-------------------------------------------------------------------------
Mass transfer across the clean, flat surface of a fully-developed turbulent open channel flow was studied in detail using large-scale direct numerical simulations at various Reynolds ($180 \leq Re_\tau \leq 630$) and Schmidt numbers ($4 \leq Sc \leq 200$).
Turbulent flow statistics were found to be in good agreement with the literature \citep[e.g.][]{Wang2019, Peruzzi2020}.  
By analysing pre-multiplied spectra and velocity fluctuations, very large scale motions (VLSM), detected previously for $Re_\tau\geq400$, were identified already for $Re_\tau\geq 365$ in the simulations  with the largest computational domain ($24H \times H \times 6H$). 
In agreement with the literature, small scale vortical structures were mainly found to be located inside the large (long-lived) low-speed streaks. 
Towards the surface, these vortices were observed to become either aligned or orthogonal to the surface.
By studying the near-surface region in detail, it was found that when approaching the surface these small scale vortices, which originate from the lower part of the turbulent boundary layer, become more uniformly distributed in the spanwise direction at high $Re_\tau$.   

Close to the surface, the vertical profiles of 
$(\overline{\langle c\rangle}-\overline{\langle c_b\rangle})/ 
( c_s-\overline{\langle c_b \rangle})$, 
${c_{rms}} / ( c_s-\overline{\langle c_b\rangle})$, and 
$j_d/j_s,j_t/j_s$ were found to collapse after normalising the $y$-coordinate using $(H-y)/\overline{\langle \delta\rangle}$. In addition, these profiles were also found to collapse with those reported in HW19, which were obtained using isotropic (rather than anisotropic) turbulence diffusing from below.  
The normalised mean transfer velocity $K_L$ was found to scale with $Sc^{-0.5}$ for Schmidt numbers up to $Sc=200$, which is in agreement with the theoretical prediction for free-slip (clean) surfaces.
The scaling of the normalised $K_L$ with Reynolds number was found to vary with $K_LSc^{-0.5} / U_b = \alpha_bRe_b^{-0.33}, K_L Sc^{-0.5} / u_\tau = \alpha_\tau Re_\tau^{-0.23}$ and $K_LSc^{-0.5} / u_\infty = \alpha_\infty Re_T^{-0.25}$. 
The two former scalings are relatively close to the data reported in literature. For $Re_T>500$, the latter scaling (with $\alpha_\infty=0.35$) is in very good agreement with HW19 and the small eddy model of \cite{Banerjee1968,Lamont1970}. Hence, the ansatz of \cite{Theofanous1976} that the scaling of $K_LSc^{-0.5} / u_\infty$ with  $Re_T^{-0.25}$ is independent on the way the turbulence is generated was found to be supported for the case of bottom-shear induced turbulence which can be replaced by isotropic turbulence diffusing from below without affecting the scaling. Note that in the open channel flow analysis $u_\infty$ and $L_\infty$ are determined at the edge of the surface influenced layer ($y=y_\infty$), which was identified as the location where  
$I(y)=
\overline{
\left( {\langle uu\rangle}+{\langle vv \rangle}+{\langle ww\rangle}  \right) / 
\left( {\langle uu\rangle}+{\langle ww\rangle} \right) }
$
 is maximum.
  
Furthermore, it was found that $K_L \sqrt{H/(D U_b)}$ scales with $0.47 \sqrt{\beta_{rms}H /U_b}$, which supports the validity of the surface divergence model for all $Re_\tau$ and $Sc$ considered.
A detailed study of the correlation $\rho(k_l,\beta)$ of the local mass transfer velocity $k_l$ and the surface divergence $\beta$ showed that it improves with increasing Reynolds number, but deteriorates with increasing $Sc$. 
Using conditional averaging, it was found that improvements in $\rho(k_l,\beta)$ with increasing Reynolds number were entirely due to increases in the correlation inside high-speed regions, while the correlation in the low-speed regions remained almost constant (and significantly higher than in the high-speed regions). 
The observed improvement in correlation was due to the surface parallel vortical structures (SPVS) - which are responsible for vertical mixing -  becoming more uniformly distributed  close to the surface with increasing Reynolds number. 

Finally, the locations of the wavenumbers $k_{\lambda_T}$, $k_\beta$  and $k_p$ associated with the Taylor micro-scale, the surface divergence integral length  and the approximate location of the peaks of the premultiplied spectral density, respectively, were determined for $Re_T=465, 1581, 2833$. It was found that $k_{\lambda_T}$ and $k_p$ almost coincide, while $k_\beta$ was significantly larger for all Reynolds numbers and well outside of the inertial range. Such that it can be concluded that the surface divergence, and hence the mean transfer velocity (which is well correlated with $\beta_{rms}$), are notably affected by the turbulence dissipation and hence the small length scales. This again supports the validity of the small eddy model for the prediction of $K_L$ for $Re_T\gtrapprox500$ in open channel flow.

%-----
\paragraph{Acknowledgements} The simulations were performed on the supercomputer bwUniCluster 2.0 supported by the state of Baden-W\"{u}rttemberg through bwHPC and on the supercomputer ForHLR II funded by the Ministry of Science, Research and the Arts Baden-W\"{u}rttemberg and by the Federal Ministry of Education and Research.

\paragraph{Funding}{We thank the Baden-W\"{u}rttemberg Stiftung gGmbH for funding M.P. (project "MOAT") through the "High Performance Computing II" program. }

\paragraph{Declaration of interests}{The authors report no conflict of interest.}

%--------
\bibliographystyle{myplainnat}
\bibliography{biblio}
%-------------------------------------------------------------------------
\end{document}